\documentclass[%
 reprint,
 amsmath,amssymb,
 aps,
floatfix,
]{revtex4-2}

\usepackage{graphicx}
\usepackage{dcolumn}
\usepackage{bm}
\usepackage{chemformula}
\usepackage{float}
\usepackage[english]{babel}

\usepackage{xcolor}
\definecolor{costumgreen}{rgb}{0, 0.91, 0.047} 
\definecolor{costumred}{rgb}{1, 0, 0.0625}
\definecolor{costumblue}{rgb}{0, 0.238, 0.918}

\usepackage[linesnumbered,ruled,vlined]{algorithm2e}
\usepackage{algpseudocode}
\usepackage{amssymb}
\usepackage{xspace}
\usepackage{caption}
\usepackage{subcaption}
\usepackage{hyperref}

\begin{document}

\preprint{APS/123-QED}

\newcommand{\newnew}[1]{{\leavevmode\color{black}#1}}

\title{The Generic Temperature Response of Large Biochemical Networks}

\author{Julian B. Voits\textsuperscript{1}}
\author{Ulrich S. Schwarz\textsuperscript{1,2}}%
 \email{Corresponding author: schwarz@thphys.uni-heidelberg.de}
\affiliation{%
\textsuperscript{1}Institute for Theoretical Physics, University of Heidelberg, Germany\\ \textsuperscript{2}BioQuant-Center for Quantitative Biology, University of Heidelberg, Germany
}%




\date{\today}

\begin{abstract}
Biological systems are remarkably susceptible to relatively small temperature changes.
The most obvious example is fever, when a modest rise in body temperature of only few Kelvin has 
strong effects on our immune system and how it fights pathogens. Another very important example
is climate change, when even smaller temperature changes lead to 
dramatic shifts in ecosystems. Although
it is generally accepted that the main effect of an increase in temperature is the acceleration of
biochemical reactions according to the Arrhenius equation, it is not clear how
it affects large biochemical networks with complicated architectures. For developmental systems like
fly and frog, it has been shown that the system response to temperature deviates
in a characteristic manner from the linear Arrhenius plot of single reactions, 
but a rigorous explanation has not been given yet. Here we use a graph-theoretical interpretation
of the mean first-passage times of a biochemical master equation to give a statistical description. 
We find that in the limit of large system size and
if the network has a bias towards a target state, then the Arrhenius plot is
generically quadratic, in excellent agreement with
numerical simulations for large networks as well as with 
experimental data for developmental times in fly and frog.
We also discuss under which conditions this generic response can be violated, for example for linear chains, which have only one spanning tree.
\end{abstract}

\maketitle


\section{\label{sec:level1}Introduction}
High fever has a dramatic effect on our body, but from the physics point
of view, it is only a modest change: increasing the human body temperature by three degrees
is less than one percent on the absolute temperature scale \cite{Blatteis}. An increase
of the same amount due to global warming would most likely result in an extensive
loss of biodiversity \cite{ipcc,garcia2018changes}. As illustrated by these examples, complex biological
systems are remarkably susceptible to changes in temperature. The explanation for
these sensitive responses to temperature was already given in the 19th century by Arrhenius, who
suggested that the rates of all chemical reactions exponentially increase with rising temperature \cite{Arrhenius}. This can be verified by an Arrhenius 
plot, in which the logarithm of the kinetic rate decreases linearly
as a function of the inverse temperature $1/T$. This insight has been confirmed
over and over again for single chemical reactions and has led to many important advances 
over the last century \cite{100yearsArrhenius}. Moreover, it has been made
more rigorous by the theories by Kramers \cite{KramersProblem,hanggi1990reaction} 
and Eyring \cite{Eyring1,Eyring2} on potential  barriers as transition states in chemical reactions. 

Biochemical systems do not behave differently from chemical systems in this regard, except
that they are typically limited by protein denaturation at high temperatures \cite{bischof2006thermal}.
A large body of experimental work exists on temperature effects in biochemical networks, ranging from 
intracellular temperature effects \cite{Begasse,Fujise,Rieder,Moore,Hayashida,Rao,GAP1,GAP2,GAP4,Evans,EColiTemperature,Vanoni,MArrest} 
to the impact of temperature on growth and development \cite{DevelopmentalArrhenius,Bliss,Powsner,BacteriaArrhenius1,BacteriaArrhenius2,BacterialGrowthT,TempDevTPyriformis,Alberghina,Yoshida,LaserCElegans}. We refer to \cite{BiologicalRatesTemperature} for a comparison of more than $1.000$ studies on biological temperature responses. 
From this body of experimental work, the picture arises that the systems response
typically does not show the Arrhenius form of single reactions. However, a fundamental
theoretical understanding of this striking observation is missing and it is an open question if such networks
show a generic response to temperature changes. Yet this question is key if one aims at understanding
biochemical systems like the immune system or ecosystems from a theoretical point of view.

There is one subject area for which the investigation of temperature effects has been
relatively systematic and comprehensive, and that is the case 
of biochemical oscillators \cite{TOs1,TOs3,TOs4,TOs5,TOs6,TOs61,TOs62,TOs63,TOs7,TOs8,TOs11, TOs2}.
Here, the general picture has emerged that they often come with compensation mechanisms which can 
assure that their functioning is unaltered within the physiological temperature range. A notable example 
for this general observation are circadian clocks, whose function is
to instruct the organism about upcoming changes due to the diurnal cycle. Because temperature changes
are one of the main consequences of changing solar input, circadian clocks are typically
temperature-compensated, to ensure reliable time measurements \cite{TOs6,TOs61,TOs62,TOs63,TOs7}. 
Another important example of temperature compensation is chemotaxis of swimming
organisms like \textit{E. Coli}, which have to reliably find food sources despite 
temperature changes and gradients in their environment \cite{EColiTemperature}.
Although the existence of temperature compensation for biochemical oscillators and chemotaxis
proves the relevance of temperature for biochemical networks, it does not instruct us about its generic effect, 
exactly because in these cases it is compensated by specific mechanisms, typically by the action
of proteins that have evolved for that purpose. 

\begin{figure*}[hbtp]
    \centering
    \includegraphics[width=0.825\textwidth]{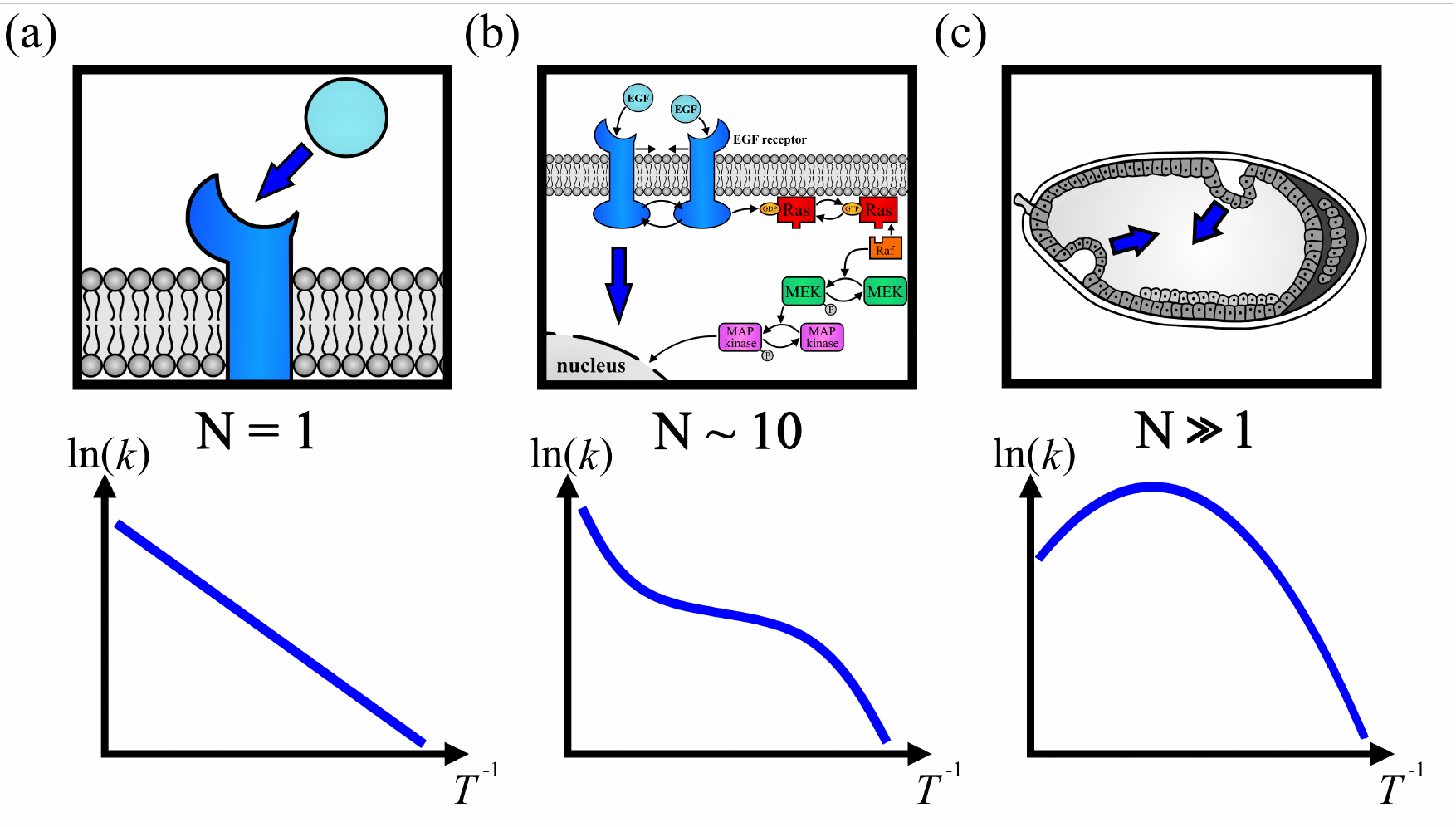}
    \caption{The Arrhenius plot depends on network size. (a) The temperature dependence of the rate of a single biochemical reaction, such as ligand-receptor binding, usually follows the Arrhenius equation, resulting in a linear relationship in the Arrhenius plot (logarithmic rates against inverse temperature). (b) For networks of intermediate size, like the MAPK-signaling pathway shown here, the temperature response is more complex and depends on microscopic details. (c) In large biochemical networks, such as those governing the development from a \textit{Drosophila} fertilized egg to a larva, a quadratic dependence emerges in the Arrhenius plot.}
    \label{fig:GeneralQuestion}
\end{figure*}

Developmental systems seem to have evolved less temperature compensation 
and therefore appear to be
better model systems to study the effect of temperature changes on large biochemical networks \cite{TonDevRates,rombouts2024mechanistic}. 
Measurements of the prepupal development time of \textit{Drosophila} under different temperature conditions were published almost one hundred years ago by Bliss \cite{Bliss}. The results clearly showed a concave curve in the Arrhenius plot, as opposed to the straight line usually found for single reactions. About a decade later, Powsner obtained similar results for the embryonic, larval and pupal phases of the development of \textit{Drosophila}, already hypothesizing that this may be a consequence of the complexity of the underlying network \cite{Powsner}. More recently, Crapse et al. published a study in which the embryonic development of \textit{Drosophila} was subdivided into twelve phases \cite{DevelopmentalArrhenius}. The authors measured the temperature dependence of the completion time of each of them. On this level, a concave temperature response emerged as well, which could be accurately described by a quadratic fit in the Arrhenius plot. In the same study, similar results were also found for \textit{Xenopus laevis} embryos.
The authors also performed simulations for a linear sequence of Poisson processes with statistically distributed parameters, but the results could not fully reproduce the shapes of the experimental curves \cite{DevelopmentalArrhenius}.
However, this approach disregards the fact that the underlying biochemical networks are bound to be
highly non-linear, with complicated feedback mechanisms and network motifs \cite{tyson2003sniffers}. \newnew{Even more recently, Rombouts et al. confirmed the quadratic shape in the Arrhenius plot of the development rates of \textit{Xenopus} embryos and also observed it for the development of zebra fish (\textit{Danio rerio}) embryos, while for \textit{Caenorhabditis elegans} the rate shows a more biphasic behavior \cite{rombouts2024mechanistic}.} In general,
although there has been some work on understanding
the temperature dependence of development rates from first principles
\cite{RatkowskyNew,Schoolfield,Schoolfield2,ModArrheniusKnies, StochasticModel}, 
the main body of work has remained empirical \cite{ DT1, DT2, DT3, DT4, DT5, DT6}.

The general question addressed in this work is illustrated in Fig.~\ref{fig:GeneralQuestion}. The temperature dependence of a single reaction is usually described well by the Arrhenius equation, resulting in a linear relationship in the Arrhenius plot (Fig.~\ref{fig:GeneralQuestion}a). However, when considering complex biological processes, e.g. the 
whole mitogen-activated protein kinase (MAPK) signaling pathway triggered by epidermal growth factors (EGF), 
that transmits extracellular signals to the nucleus and regulates differentiation \cite{MuellerEsterl}, the 
linear relation is lost and microscopic details might govern the response (Fig.~\ref{fig:GeneralQuestion}b).
In the limit of very large systems, like whole organisms such as \textit{Drosophila} developing from 
a fertilized egg to a larva \cite{muller1996developmental}, the overall rate usually approaches a quadratic dependence (Fig.~\ref{fig:GeneralQuestion}c). 
Here we provide a rigorous mathematical derivation of why large biochemical
networks typically show a quadratic shape in the Arrhenius plot. Our proof is based on
calculating first-passage times using a graph-theoretical decomposition
of the network into spanning trees and forests.
Our result is valid for all network architectures, including feedback loops, as long as several spanning trees stay relevant
in the limit of large system size.
Thus, our work solves the long-standing question of what the generic temperature response
of large and complicated biochemical networks is. To support our analytical findings,
we have developed a numerical procedure that generates networks of given sizes and
with a bias towards an endpoint, and then uses the graph-theoretical interpretation of
mean first-passage times to generate their distribution. This numerical
procedure confirms the generic temperature response predicted by our theory.
We also show that our results are in excellent agreement with 
experimental data from developmental systems like fly and frog. 
Finally we discuss the cases in which the generic response does not occur, including linear chains, which have only one spanning tree.
Our approach paves the way to
deal with smaller biochemical systems and more specific temperature responses, including
protein denaturation. 

\section{Results}
\subsection{Meaning of mean first-passage time}

A general framework to mathematically describe a chemical reaction network is to express it as a time-dependent probability distribution $p_i(t)$ on $N_v$ discrete states, where without loss of generality we denote states $1$ and $N_v$
as start and target states, respectively. For a linear chain, we would have $N = N_v-1$ reactions, each with
forward and backward directions, while in general, the system could contain many loops. In the following, we will first use $N$ as
main variable; for the numerical part coming later,
it is more convenient to use $N_v = N+1$.

The time evolution of $p_i$ is governed by the master equation \cite{ChemMEDerivation, schnoerr2017approximation}:
\begin{equation}
    \dot{p}_i(t)=\sum_{j=1}^{N+1} \Big (k_{ji}p_j(t)-k_{ij}p_i(t)\Big ),\label{eq:MasterEquation}
\end{equation}
or in vectorial notation:
\begin{equation}
    \dot{\vec{p}}_i(t)=K\vec{p}(t),\label{eq:MEvectorial}
\end{equation}
where $k_{ij}$ is the transition rate from state $i$ to state $j$ and
\begin{align}
    K_{ij}=\begin{cases}\hspace{0.6cm} k_{ji} &\text{if }i\neq j\\ -\sum_m k_{im} &\text{if }i= j\end{cases}\ .
\end{align}
Realistic biochemical networks typically exhibit a large number of intertwined reactions, meaning that $N \gg 1$ and Eq. (\ref{eq:MasterEquation}) then describes a large system of coupled ordinary differential equations,
for which it is difficult to find a full mathematical solution. Another important limitation is that 
often not all connections and/or rates might be known. Together, this raises the question if one can make
progress without explicitly solving the complete system.

Indeed, certain features of the system described by Eq. (\ref{eq:MasterEquation}) can be computed without the need for a complete solution \cite{nam2022linear}. A notable example are first-passage times (also called first-hitting times or exit times) \cite{Redner,vanKampen} and in particular, the mean first-passage time (MFPT), which characterizes the typical completion time of the process of interest. The main example in this work is a developmental process seen as the consequence of a complex network of biochemical reactions. Developmental
systems are very large and complex and besides biochemistry also involve many spatial processes, such as cytoskeletal
rearrangements \cite{kanesaki2011dynamic}. However, experimental measurements of heat generation in zebrafish embryos 
combined with modeling of biochemical networks
suggest that even in this case, the rate-limiting steps are in the biochemical control system for the
cell cycle \cite{rodenfels2019heat}, which in itself is again a large and complex biochemical network \cite{ferrell2011modeling}.
To give an instructive example of a transparent biochemical network that is active during development
and that can be described with the master equation from Eq. (\ref{eq:MasterEquation}), one can think of the mitogen-activated
protein kinase (MAPK) signaling pathway triggered by epidermal growth factors (EGF) shown in Fig. \ref{fig:GeneralQuestion}b. In this case, the MFPT $\langle\tau\rangle$ describes the typical time for the signal to reach from the plasma membrane to the nucleus, where gene expression is changed. Here we ask how such MFPTs can be calculated for this and more complex networks, and how they depend on temperature.

\subsection{Solution with graph theory}

\subsubsection{Some concepts from graph theory}

It is often a useful approach to represent networks described by a master equation as weighted and directed graphs, i.e., vertices and directed edges (pairs of vertices), where the edges have a weight attributed to them \cite{lyons2017probability}. Then the vertices are the states, the directed edges are the possible jumps and their weights are the jumping rates $k_{ij}$ from vertex $i$ to vertex $j$. In the following, graphs are always understood as weighted and directed. Also, the rates $k_{ij}$ are simply referred to as edges, for the ease of notation, setting $k_{ij}=0$ if there is no transition from $i$ to $j$.

\begin{figure}[h]
    \vspace{0.5cm}
    \centering
    \includegraphics[width=0.48\textwidth]{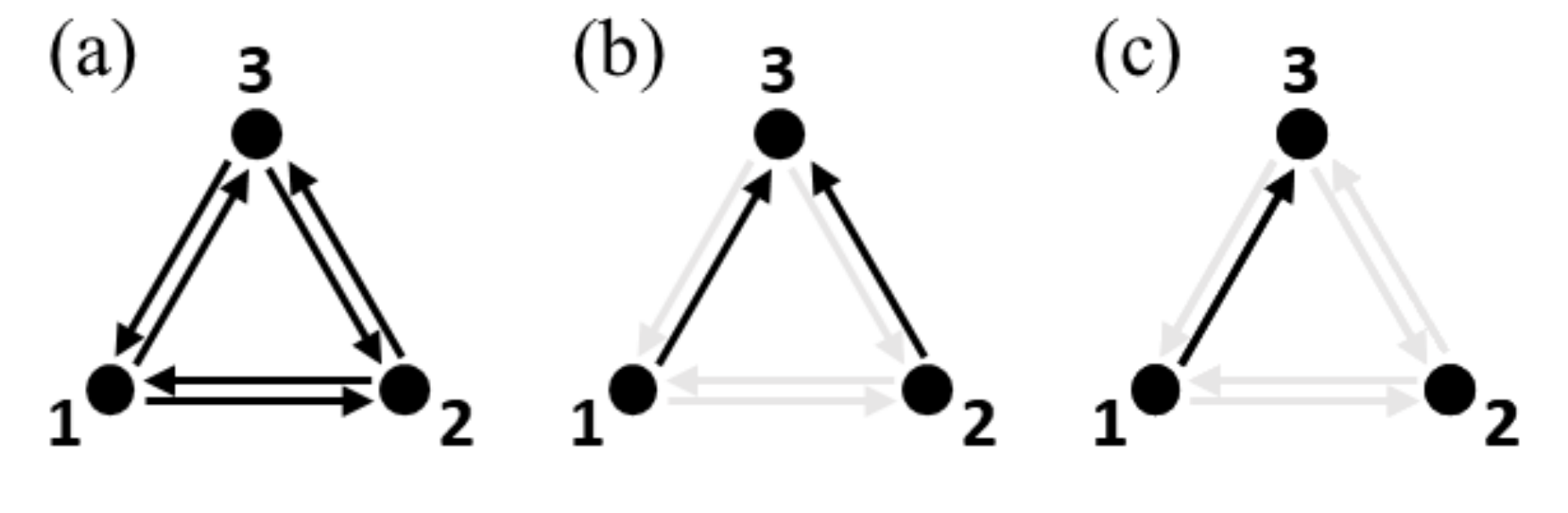}
     \caption{An illustration of the graph-theoretical decomposition of networks into spanning trees and forests used throughout this work. (a) The full directed graph on three vertices. (b) A spanning tree rooted at $3$. (c) A spanning forest of two trees rooted at $2$ and $3$.}
     \label{fig:K3Example}
\end{figure}

A tree rooted at a vertex $i$ is a cycle-free graph where there is a directed path from any other vertex to $i$. Note that this requires $i$ to be an endpoint because any outgoing edge from $i$ would create a cycle. It is then also clear that the root of a tree is unique. A disjoint union of trees is called a forest.
A spanning tree (forest) is a subgraph of a given graph that contains all its vertices and is a tree (forest). This is illustrated in Fig. \ref{fig:K3Example}.

The (in-degree) Laplacian matrix $L$ of a graph is defined as:
\begin{equation}
    L_{ij}:=\begin{cases} \hspace{0.33cm}-k_{ij}&\text{if }i\neq j\\ \sum_{m\neq i} k_{im} &\text{if }i= j\end{cases},
\end{equation}
so the $i$-th entry on the main diagonal is the sum of all incoming edges and $L_{ij}$ for $i\neq j$ are the negative transition rates from $i$ to $j$.    
\subsubsection{Graph-theoretical interpretation of FPT moments}
The standard textbook approach to find first-passage times (FPTs) is to evaluate them in Laplace space \cite{Honerkamp,vanKampen,Redner}. However, this becomes increasingly cumbersome for larger
systems. Here, we instead use a solution that can be derived with graph theory \cite{NamPhD,FPTGunawardena,khodabandehlou2022trees}.
Consider the master equation on a finite space $i=1,...N+1$ with time-independent rates 
$k_{ij}$. Then the formal solution to Eq. (\ref{eq:MEvectorial}) is given by:
\begin{equation}
    \vec{p}(t)=\text{exp}(Kt)\vec{p}_0,\label{eq:HomogeneousMEsolution}
\end{equation}
with $\text{exp}$ denoting the matrix exponential and $\vec{p}_0=\vec{p}(0)$. Since $1-p_{N+1}=\sum_{i=1}^{N}p_i$, Eq. (\ref{eq:HomogeneousMEsolution}) can be restricted to only the first $N$ components. In particular, $K$ is then no longer singular.

Asking for the FPT to reach state $N+1$, this state can be assumed to be absorbing, i.e., $k_{N+1i}=0$ for all $i$, 
because we are only interested in the first time that it is reached. 
The FPT density $f(t)$ then follows from:
\begin{align}
    p(\tau\leq t) &= p_{N+1}(t)\\ \implies\hspace{0.6cm}  f(t)&=\dot{p}_{N+1}(t)\nonumber\\ &= \hat{e}^T_{N+1} K \vec{p}(t)\nonumber\\&=\hat{e}_{N+1}^T K\text{exp}(Kt)\vec{p}_0,
\end{align}
where $\hat{e}_i$ denotes the i-th unit vector. We define the FPT density to reach $N+1$ starting at state $i$
\begin{align}
    f_i(t):&=\hat{e}_{N+1}^T K\text{exp}(Kt)\hat{e}_i\nonumber \\&= \hat{e}_i^T \text{exp}(K^Tt)K^T\hat{e}_{N+1},
\end{align}
where the second equality uses the symmetry of the inner product. In vector form, this now reads:
\begin{align}
    \vec{f}&=\text{exp}(K^Tt)K^T\hat{e}_{N+1}\nonumber\\ &=\text{exp}(K^Tt)\vec{f}_0,\label{eq:GeneralSolutionFPTDensityME}
\end{align}
where $\vec{f}_0= (k_{1,N+1},...k_{N,N+1})^T$. A distribution of this shape is called phase-type distribution \cite{MatrixExpDistributions}. The moments are obtained by integration by parts:
\begin{align}
    \langle\vec{\tau ^n}\rangle &=\int \text{d}t\hspace{0.1cm} t^n \vec{f} \nonumber\\&=-n\big (K^{T}\big )^{-1}\int \text{d}t\hspace{0.1cm}t^{n-1} \text{exp}(K^Tt)\vec{f}_0\nonumber \\&=-n\big (K^{T}\big )^{-1}\langle\vec{\tau ^{n-1}}\rangle.
\end{align}
Equivalently, we have
\begin{equation}
    K^T\langle\vec{\tau ^n}\rangle=-n\langle\vec{\tau ^{n-1}}\rangle.\label{eq:RecursiveRelFPTMoments}
\end{equation}
Here $\langle\tau ^n_i\rangle$ is the $n-$th moment of the FPT when starting at state $i$.
Iterative application of the formula above yields:
\begin{equation}
    \langle\vec{\tau ^n}\rangle=n!\big ((-K^{T})^{-1}\big )^n\vec{1}\label{eq:FPTMoments},
\end{equation}
where $\vec{1}=(1,1,...,1)^T$. Note that
\begin{equation}
    (-K^{T})_{ij=1,...N}=\begin{cases} \hspace{0.33cm}-k_{ij}&\text{if }i\neq j\\ \sum_{m=1}^{N+1} k_{im} &\text{if }i= j\end{cases}
\end{equation}
can be interpreted as the submatrix obtained by deleting the $N+1$-th row and column of the Laplace matrix of the weighted and directed graph identifying the states with vertices and the edges with the transitions between them, where $k_{ij}$ is the weight for the edge going from $i$ to $j$.

\begin{figure}[h]
    \centering
    \includegraphics[width=0.4\textwidth]{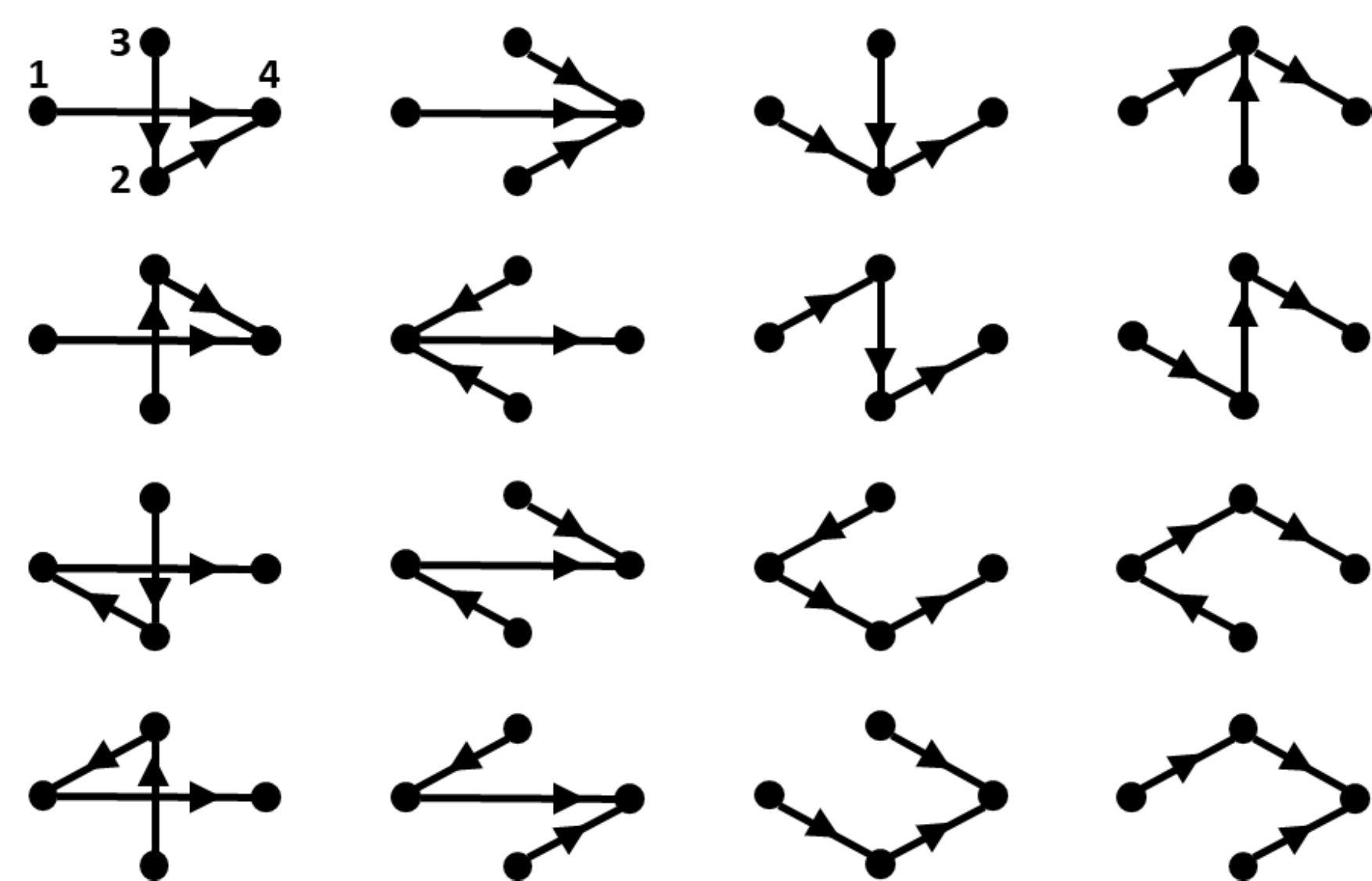}
    \caption{All spanning trees of the complete graph on four vertices rooted at $4$. The graph-theoretical interpretation of the sum in Eq. \ref{eq:sum_spanning_trees} consists in finding the spanning trees. The vertex labels are for all graphs as indicated for the first one.}
    \label{fig:TreesK4}
\end{figure}

By a generalization of Kirchhoff's theorem to weighted and directed graphs \cite{GeneralizedKirchhoff},
the determinant of $-K^{T}$ is given by:
\begin{equation}
    \det{\Big (-K^{T}\Big )}=\sum_{\mathcal{T}_{[N+1]}}w(\mathcal{T})\label{eq:sum_spanning_trees}, 
\end{equation}
where $\mathcal{T}_{[N+1]}$ denotes the spanning trees of the corresponding graph rooted at $N+1$ and their weights are defined as $w(\mathcal{T}):=\prod_{k_{ij}\in E(\mathcal{T})}k_{ij}$, with $E(\mathcal{T})$ being the edge set of $\mathcal{T}$. Fig. \ref{fig:TreesK4} shows $\mathcal{T}_{[4]}$ for the complete graph on four vertices as an example.

\begin{figure}[h]
    \centering
    \includegraphics[width=0.4\textwidth]{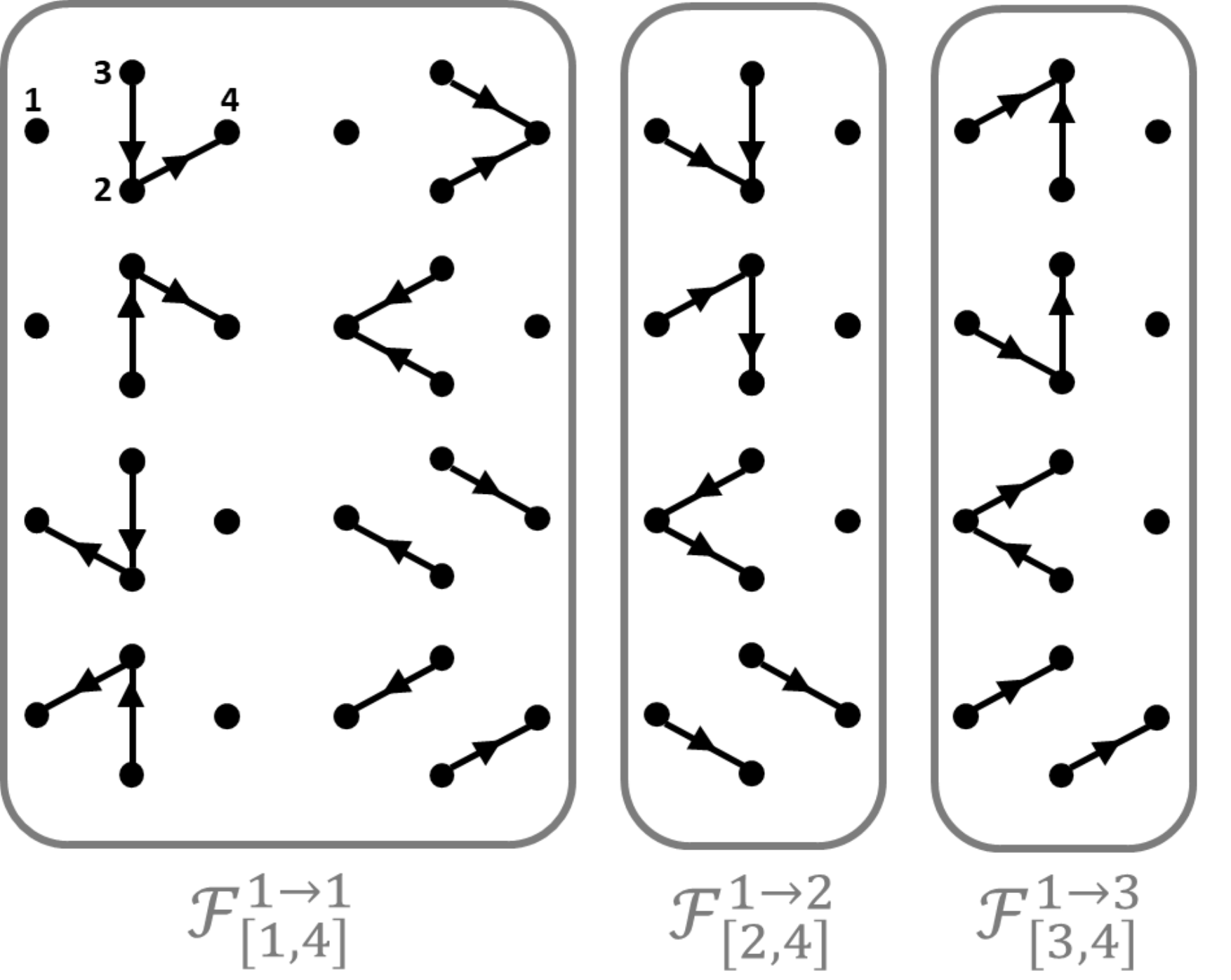}
    \caption{All spanning forests of the complete graph on four vertices rooted at $4$. The graph-theoretical interpretation of the summands in Eq.\ref{eq:sum_spanning_forests} consists in finding the two-tree spanning forests $\mathcal{F}_{[1,4]}^{1\rightarrow 1}$, $\mathcal{F}_{[2,4]}^{1\rightarrow 2}$ and $\mathcal{F}_{[3,4]}^{1\rightarrow 3}$ of the complete graph on four vertices rooted at $4$. The vertex labels are for all graphs as indicated for the first one.}
    \label{fig:ForestsK4}
\end{figure}

$(-K^{T})^{-1}$ can be expressed as:
\begin{equation}
    (-K^{T})^{-1}_{ij}=\frac{(-1)^{i+j}}{\det{(-K^{T}})}M_{ji},
\end{equation}
where $M_{ji}$ is the determinant of the $(j,i)$-minor of $-K^{T}$, which can also be expressed as a sum with a graph-theoretical interpretation \cite{NamPhD}:
\begin{equation}
    M_{ji}=(-1)^{i+j}\sum _{\mathcal{F}_{[j,N+1]}^{i\rightarrow j}}w(\mathcal{F}),\label{eq:sum_spanning_forests}
\end{equation}
where $\mathcal{F}_{[j,N+1]}^{i\rightarrow j}$ are the spanning forests with two trees of the graph, one rooted at $j$ and containing $i$ and the other one rooted at $N+1$.  Fig. \ref{fig:ForestsK4} shows $\mathcal{F}_{[i,4]}^{1\rightarrow i}$ for the complete graph on 4 vertices.

This yields
\begin{equation}
    (-K^{T})^{-1}_{ij}=\frac{\sum _{\mathcal{F}_{[j,N+1]}^{i\rightarrow j}}w(\mathcal{F})}{\sum_{\mathcal{T}_{[N+1]}}w(\mathcal{T})},\label{eq:ComponentsInverseLaplacianGraphTheory}
\end{equation}
and therefore, Eq.(\ref{eq:FPTMoments}) becomes \cite{NamPhD}:

\begin{equation}
    \langle\vec{\tau ^n}\rangle=n!\sum_{j_0,...,j_n=1}^N\frac{\prod_{m=1}^{n}\sum _{\mathcal{F}_{[j_m,N+1]}^{j_{m-1}\rightarrow j_{m}}}w(\mathcal{F})}{\big (\sum_{\mathcal{T}_{[N+1]}}w(\mathcal{T})\big )^n}\hat{e}_{j_0}.\label{eq:FPTMomentsGraphTheory}
\end{equation}

The mean in particular reads \cite{NamPhD,FPTGunawardena}:
\begin{equation}
    \langle\tau\rangle =\frac{\sum_{j=1}^N\sum _{\mathcal{F}_{[j,N+1]}^{1\rightarrow j}}w(\mathcal{F})}{\sum_{\mathcal{T}_{[N+1]}}w(\mathcal{T})}.\label{eq:MFPTFormula}
\end{equation}
Note that similar graph-theoretical counting schemes also exist for the steady state distributions of ergodic networks \cite{MarkovChainTheoremHill,Zia}. In Appendix \ref{app:MFPTCountingExample}, we show how
this formula can be used to derive MFPTs for two example networks for which the results are also
known from Laplace transforms.

Overall, the graph-theoretical approach results in three possible approaches to the FPTs of a homogeneous master equation. 
The first one consists in solving the matrix exponential as given in Eq. (\ref{eq:GeneralSolutionFPTDensityME}) or, alternatively, the corresponding differential equation:
\begin{equation}
    \dot{\Vec{f}}=K^T\Vec{f},
\end{equation}
with $\vec{f}_0$ as above. This is numerically feasible for systems of moderate size.
If no full analytic solution is available, the moments of the FPT can be computed by either inverting $-K^T$ algebraically and using Eq. (\ref{eq:FPTMoments}) or by solving the combinatorial problem in Eq. (\ref{eq:FPTMomentsGraphTheory}).

First-passage time distributions can be broad, especially when considering the statistics of rare events, and higher moments are therefore often of interest \cite{Redner}. For processes with a clear bias towards a final state, the distribution of the completion time tends to be more narrow and its mean characterizes it sufficiently well \cite{SimplicityComplexNW}, which is likely to be the case for most developmental systems. Indeed, the development times of \textit{Drosophila} embryos vary only by a few percent \cite{LowCV}.  

\subsection{Effect of temperature}

Temperature now enters the description via the rate constants. By the Arrhenius equation \cite{Arrhenius}, the temperature dependence of the $k_{ij}$ can be described as follows:
\begin{align}
    k_{ij}&=A_{ij}e^{-\frac{E_{ij}}{k_BT}}=A_{ij}e^{-\frac {E_{ij}[\frac{\text{J}}{\text{mol}}]}{RT}},
\end{align}
where $T$ denotes the temperature, $k_B$ Boltzmann's constant, $A_{ij}$ a (temperature-independent) prefactor and $E_{ij}$ is the activation energy. For large systems like developmental ones, one often
expresses $E_{ij}$ in $\frac{\text{J}}{\text{mol}}$. Then $k_B$ has to be replaced by the universal gas constant $R$.

Since most biological organisms have evolutionarily adapted to a relatively fixed thermal environment, it is reasonable to rephrase the Arrhenius equation in terms of the rate constant at a reference temperature $T_0$, introducing
$k^0_{ij}=k_{ij}(T=T_0)$. $T_0$ describes the temperature at which the organism usually operates. For instance, for endothermic animals, the body temperature would serve as the natural reference point, while for exothermic animals, it would be the typical temperature of the environment. The $k^0_{ij}$ can be seen as the standard rates, and a crucial assumption for the following analysis is that these standard rates are independent of the activation energies $E_{ij}$. The Arrhenius equation can then be expressed in terms of these parameters as:
\begin{align}
    k_{ij}&=A_{ij}e^{-\frac{E_{ij}}{k_BT}}=k^0_{ij} e^{-\Delta\beta E_{ij}} \label{eq:AltArrhenius}
\end{align}
where $\Delta\beta := \frac{1}{k_BT}-\frac{1}{k_BT_0}$.

\subsection{Partition sums and generating functions}

We note that the number of spanning trees on a complete graph on $N+1$ vertices is given by the Cayley 
formula \cite{Cayley} as $(N+1)^{N-1}$, so it grows faster than exponentially with $N$. Even though many of the rates vanish, for sufficiently complex networks one can expect the numerator and the denominator of Eq. (\ref{eq:MFPTFormula}) to be the sum of many such graphs. This motivates the introduction of a partition sum-like quantity $Z$ summing over the spanning trees:
\begin{equation}
    Z_{\mathcal{T}}=\sum_{\mathcal{T}_{[N+1]}}w(\mathcal{T})\label{eq:PartitionSumTrees},
\end{equation}
and for the two-tree spanning forests: 
\begin{equation}
    Z_{\mathcal{F}}=\sum_{j=1}^N\sum _{\mathcal{F}_{[j,N+1]}^{1\rightarrow j}}w(\mathcal{F})\label{eq:PartitionSumReducedTrees},
\end{equation}
so that Eq. (\ref{eq:MFPTFormula}) reads in terms of these two quantities:
\begin{equation}
    \langle\tau\rangle =\frac{Z_{\mathcal{F}}}{Z_\mathcal{T}}.\label{eq:MFPTRatioOfZs}
\end{equation}
Letting the sums in Eq. (\ref{eq:PartitionSumTrees}) and Eq. (\ref{eq:PartitionSumReducedTrees}) run only over the non-vanishing trees, all involved rates satisfy $k_{ij}>0$, and one can parameterize them as $k_{ij}(a)=e^{aX_{ij}}$ such that $k_{ij}(a=1)=e^{X_{ij}}$ is the rate of interest. Applying this parametrization, one obtains for $Z_\mathcal{T}$:
\begin{align}
    Z_\mathcal{T}(a)&=\sum_{\mathcal{T}_{[N+1]}} \prod_{k_{ij}\in E(\mathcal{T})}k_{ij}\nonumber\\ &=\sum_{\mathcal{T}_{[N+1]}}e^{a\sum_{k_{ij}\in E(\mathcal{T})} X_{ij}}\nonumber\\ &=\sum_{\mathcal{T}_{[N+1]}}e^{a X_{\mathcal{T}}}\label{eq:ZaAnsatz},
\end{align}
defining $X_{\mathcal{T}}:=\sum_{k_{ij}\in E(\mathcal{T})} X_{ij}$, which is the sum over all $X_{ij}$ of a given tree. Now consider the Taylor series of $\ln{Z_\mathcal{T}}$ around $a=0$:
\begin{align}
    \ln{Z_\mathcal{T}}(a)=\sum_{n=0}^\infty \frac{\partial ^n  \ln{Z_\mathcal{T}}}{\partial^n a}\Big |_{a=0}\frac{a^n}{n!}\ .
\end{align}
The zeroth order coefficient is 
\begin{equation}
    \ln{Z_\mathcal{T}}(a=0)=\ln{\Big (\sum_{\mathcal{T}_{[N+1]}}1 \Big )}=\ln{|\mathcal{T}|},
\end{equation}
i.e., it is the logarithm of the total number of spanning trees. For the first order, one finds:
\begin{align}
    \frac{\partial  \ln{Z_\mathcal{T}}}{\partial a}\Big |_{a=0}=\frac{\sum_{\mathcal{T}_{[N+1]}}X_{\mathcal{T}}}{|\mathcal{T}|}=\langle X_\mathcal{T} \rangle _\mathcal{T},
\end{align}
i.e., the mean of $X_{\mathcal{T}}$ over the trees. For the second order, one obtains:
\begin{align}
    \frac{\partial ^2 \ln{Z_\mathcal{T}}}{\partial a^2}\Big |_{a=0}&=\frac{\sum_{\mathcal{T}_{[N+1]}}X_{\mathcal{T}}^2}{|\mathcal{T}|}-\Big(\frac{\sum_{\mathcal{T}_{[N+1]}}X_{\mathcal{T}}}{|\mathcal{T}|}\Big )^2\nonumber\\ &=\langle X_\mathcal{T}^2 \rangle _\mathcal{T}-\langle X_\mathcal{T} \rangle _\mathcal{T}^2\nonumber\\ &=\sigma_{X_\mathcal{T}}^2,
\end{align}
i.e., the variance of $X_\mathcal{T}$ over the trees. These results are no coincidence: Eq. (\ref{eq:ZaAnsatz}) is precisely the cumulant generating function for $X_\mathcal{T}$ such that one finds in general:
\begin{align}
    \frac{\partial ^n \ln{Z_\mathcal{T}}}{\partial a^n}\Big |_{a=0}=\kappa_n^{X_\mathcal{T}}.
\end{align}
Hence, one can express $\ln{Z_\mathcal{T}}$ as:
\begin{align}
    \ln{Z_\mathcal{T}(a)}=\sum_{n=0}^\infty \kappa_n^{X_\mathcal{T}}\frac{a^n}{n!},
\end{align}
with the convention that $\kappa _0^{X_\mathcal{T}}:=\ln{|\mathcal{T}|}$. Completely analogously for $\ln{Z_{\mathcal{F}}}$ one gets:
\begin{align}
    \ln{Z_{\mathcal{F}}(a)}=\sum_{n=0}^\infty \kappa_n^{X_\mathcal{F}}\frac{a^n}{n!},
\end{align}
with $\kappa_0^{X_\mathcal{F}}:=\ln{|\mathcal{F}|}$.
For the rates of interest ($a=1$), the logarithm of Eq. (\ref{eq:MFPTRatioOfZs}) becomes:
\begin{align}
    \ln{\langle\tau\rangle}&=\ln{Z_{\mathcal{F}}(a=1)}-\ln{Z_{\mathcal{T}}(a=1)}\nonumber\\ &=\sum_{n=0}^\infty \frac{\kappa_n^{X_\mathcal{F}}-\kappa_n^{X_\mathcal{T}}}{n!}\label{eq:GeneralLogMFPT}.
\end{align}
Note that Eq. (\ref{eq:GeneralLogMFPT}) is completely general
and, as such, could be applied, for example, to the MAPK-system
shown in Fig. \ref{fig:GeneralQuestion}b. 
The cumulants are taken over the spanning trees and the expression is precise if all $k_{ij}$ are known for any $N$. 
Also, our treatment so far is not restricted to biochemical networks. In principle, it can also be applied to other settings described by a master equation. 

\subsection{Limit of large system size}
\label{sec:LargeSystemLimit}
We now are in a position to study the limit 
of large networks ($N \gg 1$). Because in this case
not all details can be known, at the same time
we switch to a statistical description for the 
reaction rates $k_{ij}$. To describe the effect of temperature, we consider biochemical networks with rates that have an Arrhenius-like temperature dependence, that is to say, $X_{ij}=-\Delta\beta E_{ij}+\ln{k^0_{ij}}$, where $E_{ij}$ and $\ln{k^0_{ij}}$ are assumed to be independent random variables. For that case, using that 
\begin{align}
   \kappa_n^{X_\mathcal{T}}&=\kappa_n^{-\Delta\beta E_{\mathcal{T}}+\sum_{k_{ij}\in E(\mathcal{T})} \ln{k^0_{ij}}}\nonumber\\ &=(-\Delta\beta)^n\kappa_n^{E_{\mathcal{T}}}+\kappa_n^{\sum_{k_{ij}\in E(\mathcal{T})}\ln{k^0_{ij}}},
\end{align}
where $E_\mathcal{T}:=\sum_{k_{ij}\in E(\mathcal{T})} E_{ij}$, one obtains:
\begin{align}
    \ln{\langle\tau\rangle}=\sum_{n=1}^\infty \frac{(-1)^n}{n!}(\kappa_n^{E_\mathcal{F}}-\kappa_n^{E_\mathcal{T}})\Delta\beta^n+const.\label{eq:GlobalArrheniusGeneralFormula}
\end{align}

Note that backward rates do not fully enter in the spanning trees (cf. examples in Appendix \ref{app:MFPTCountingExample}), so the activation energies on the trees and on the two-tree forests are different in general. The result so far requires to know the cumulants $\kappa_n^{E_\mathcal{T}}$ and $\kappa_n^{E_\mathcal{F}}$ to get the global rate constant. 

If now $E_\mathcal{T}$ and $E_\mathcal{F}$ can be sufficiently well approximated by normal distributions, $E_\mathcal{T}\sim \mathcal{N}(\langle E\rangle _\mathcal{T},\sigma_\mathcal{T}^2)$ and $ E _\mathcal{F}\sim \mathcal{N}(\langle E\rangle _\mathcal{F},\sigma_\mathcal{F}^{2})$, then all but the first two cumulants vanish, resulting in the following expression:
\begin{align}
    \ln{\langle\tau\rangle}=(\langle E\rangle _\mathcal{T}-\langle E\rangle _\mathcal{F})\Delta\beta +\frac{\sigma _\mathcal{F}^{2}-\sigma_\mathcal{T} ^{2}}{2}\Delta\beta ^2 +const.\label{eq:GeneralQuadraticResponse}
\end{align}
yielding a quadratic dependence in the Arrhenius plot (since $\ln{\langle \tau\rangle}=-\ln{k}$). 

$E_\mathcal{T}=\sum_{k_{ij}\in E(\mathcal{T})} E_{ij}$ and $E_\mathcal{F}=\sum_{k_{ij}\in E(\mathcal{F})} E_{ij}$ are the sums of $N$ and $N-1$ independent random variables, respectively. Since $N \gg 1$, the central limit theorem suggests that this is a valid assumption as long as the higher cumulants vanish fast enough compared to the differences of the first two terms, requiring sufficiently well-behaved convergence to the normal distribution.
\newnew{The simplest case is that the $E_{ij}$ already follow a normal distribution.
However, any distribution, including the uniform one, is sufficient, 
as long as it satisfies the conditions required for the central limit theorem.}
Then one obtains:
\begin{align}
    \ln{\langle\tau\rangle}=&(N\langle E_{ij}\rangle _\mathcal{T}-(N-1)\langle E_{ij}\rangle _\mathcal{F})\Delta\beta \nonumber\\ &+\frac{(N-1)\sigma _{\mathcal{F},E_{ij}}^{2}-N\sigma_{\mathcal{T},E_{ij}} ^{2}}{2}\Delta\beta ^2 +const.
\end{align}
This constitutes our main theory result: the Arrhenius plot for large and complex biochemical
networks is expected to show a quadratic dependence on the inverse temperature.

\subsection{Comparison to simulated networks}

\begin{figure}[h]
     \centering
    \includegraphics[width=0.4\textwidth]{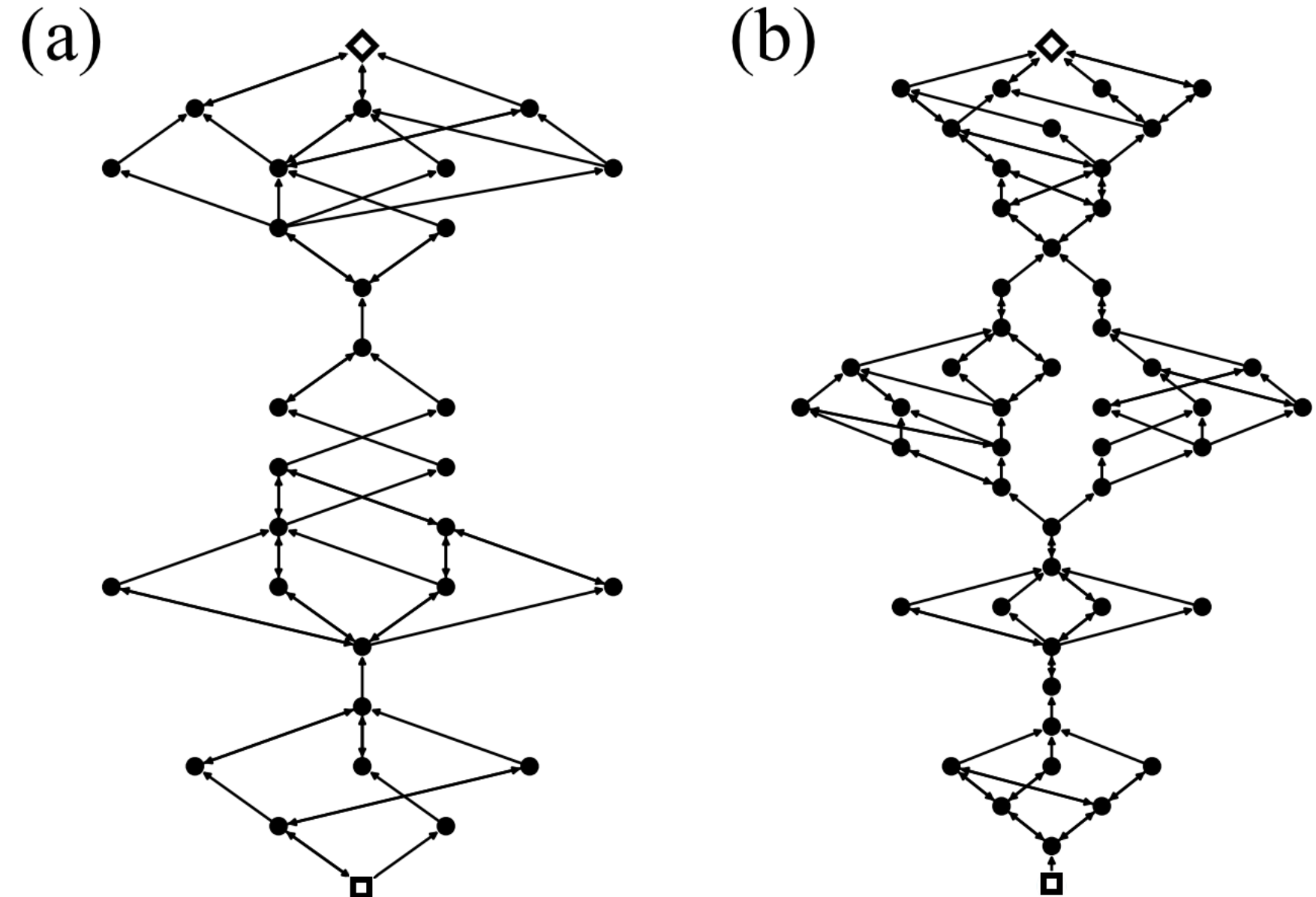}
     \caption{Randomly generated networks with features
     that are typical for development of sizes (a) $N_v=30$
     and (b) $N_v=50$ vertices. The initial vertices are represented as squares and the final vertices as diamonds. The networks are
     biased towards the final vertices and contain compact
     elements with many connections.}
     \label{fig:Random_Graphs}
\end{figure}

To illustrate the convergence predicted by our theory, random networks on $N_v = 10$ to $N_v = 1.000$ vertices displaying some basic characteristics of the biochemical network of a developmental process were generated  (see Appendix \ref{app:DetailsSimulation} for details) \cite{Github2025}. In particular, they were designed to have a relatively large distance between the initial vertex $0$ and the final vertex $N_v = N+1$ as well as a block-like structures with many connections and irreversible transitions between them, reflecting the checkpoints present in development. Two small networks with $N_v=30$ and $N_v=50$ are shown as examples in Fig. \ref{fig:Random_Graphs}a and b, respectively. 

\begin{figure}[h]
     \centering
    \includegraphics[width=0.5\textwidth]{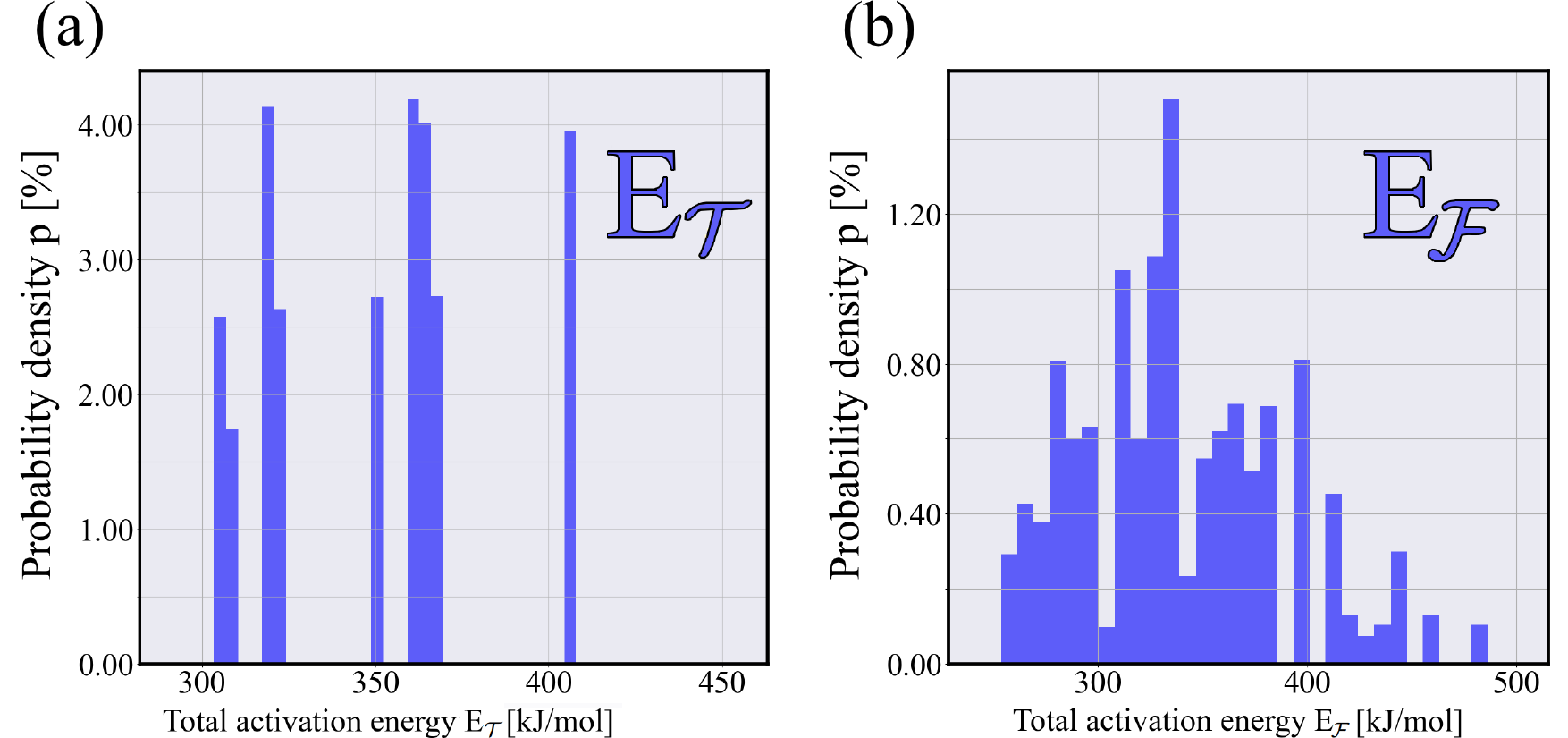}
     \caption{Histograms of the total activation energies of (a) spanning trees $E_{\mathcal{T}}$ and (b) two-tree spanning forests $E_{\mathcal{F}}$, i.e., the sum of all the activation energies along their edges, on randomly generated graphs on $N_v=10$ vertices.}
     \label{fig:Histogram_10}
\end{figure}
\begin{figure*}[hb]
     \centering
    \includegraphics[width=1\textwidth]{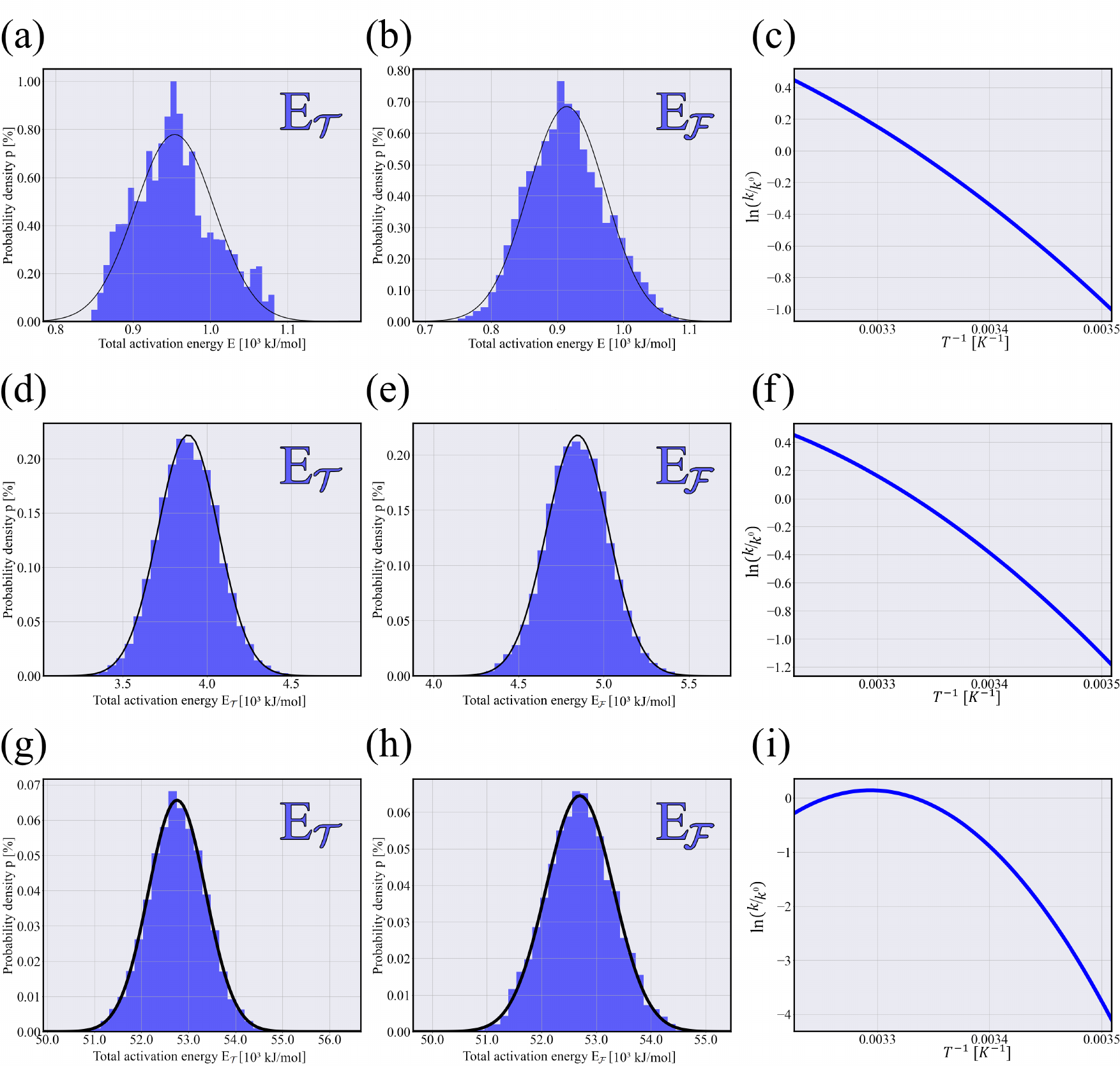}
     \caption{Histograms and Arrhenius plots for simulated random networks of increasing sizes. (a-c) $N_v =20$. (d-f) $N_v=100$. (g-i) $N_v=1.000$ (a,d,g) Spanning trees. (b,e,h) Spanning forests. (c,f,i) Resulting Arrhenius plots. The fitted parameters are listed in Appendix \ref{app:FitParameters}.}
     \label{fig:Histograms_and_Arrhenius}
\end{figure*}
The spanning trees and two-tree spanning forests of these base networks were then drawn at random. A standard procedure to generate random spanning trees is given by Wilson's algorithm \cite{wilson1996generating}. While removing a random edge from a spanning tree will result in a spanning forest, this procedure does not yield all spanning forests with the same probability. In order to ensure an unbiased sample, a specific algorithm was devised for the generation of two-tree spanning forests. In Appendix \ref{app:Algorithm}, the steps needed for this algorithm are explained and its functionality is illustrated for a simple example network. 

The activation energies were assigned randomly to the edges. The activation energies were drawn at random from the intervals $[60 \frac{\text{kJ}}{\text{mol}},100 \frac{\text{kJ}}{\text{mol}}]$ for the forward rates and from $[20 \frac{\text{kJ}}{\text{mol}},60 \frac{\text{kJ}}{\text{mol}}]$ for the backward rates. These represent typical values of activation energies of biochemical reactions listed in the literature\cite{DevelopmentalArrheniusRanges}. 

In each case, $10.000$ spanning trees and at least $10.000$ two-tree spanning forests  (note that the number of generated forests is not predefined to ensure an unbiased sample, as explained in Appendix \ref{app:Algorithm}) and the total activation energies for each of them were calculated. The total activation energies for $N_v=10$ are shown in Fig. \ref{fig:Histogram_10} as histograms. One observes that there are just a few spanning trees and also a moderate number of two-tree spanning forests, so that the distribution is not reminiscent of a Gaussian yet. The other histograms, along with fits to normal distributions, are shown in Fig. \ref{fig:Histograms_and_Arrhenius}. As expected, the histograms of the total activation energies approach a normal distribution as $N_v$ increases. 

\begin{figure}[b]
     \centering
    \includegraphics[width=0.5\textwidth]{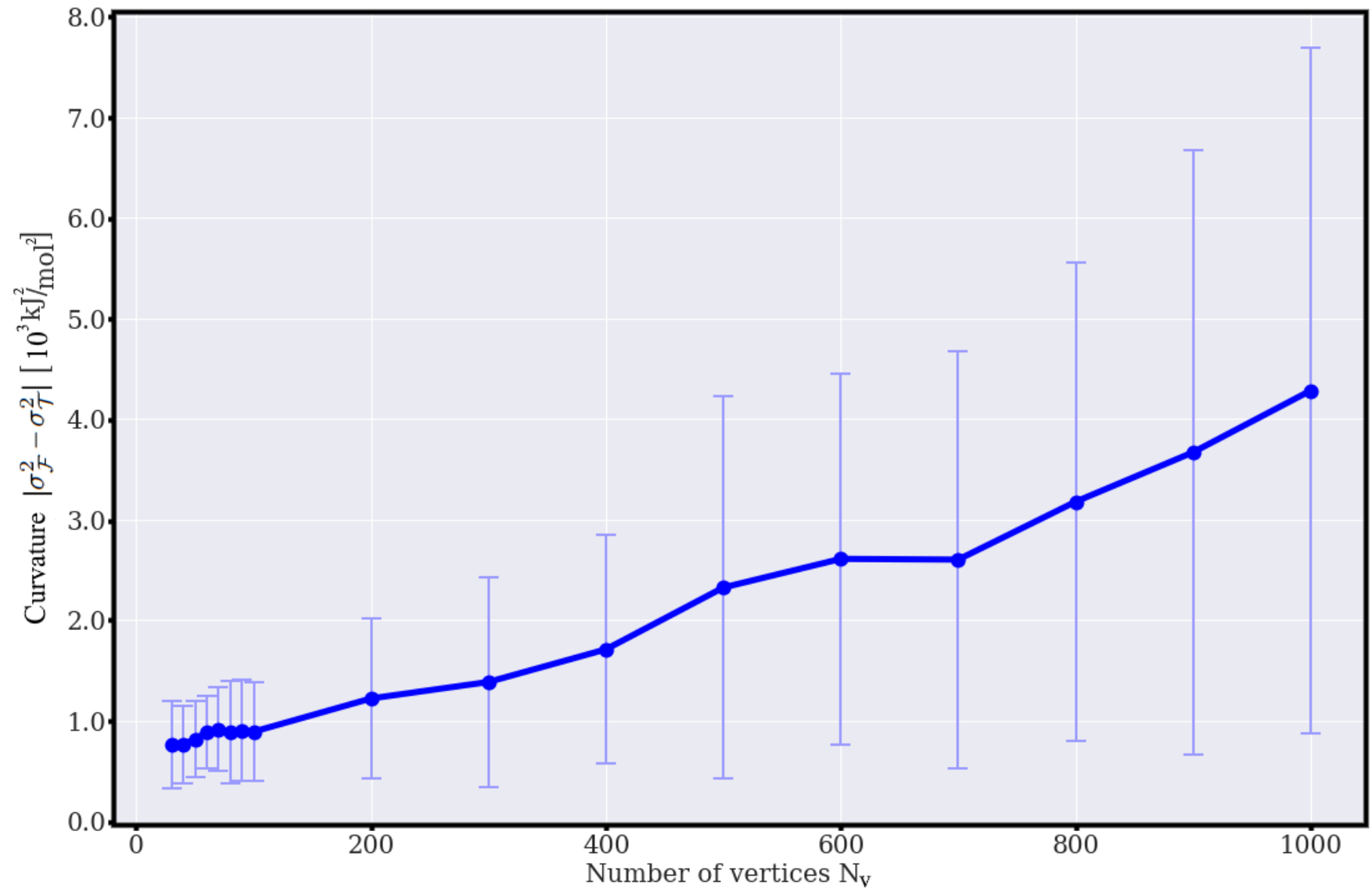}
     \caption{The absolute value of the quadratic term in the Arrhenius plot $\sigma _F^2-\sigma _T^2$ versus number of vertices $N_v$. For every value $N_v$, 100 networks of this size were generated and the distribution of the total activation energies of the trees and two tree spanning forests were computed. The error bars indicate the standard deviations.}
     \label{fig:quadratic_vs_N}
\end{figure}

Then, Eq. (\ref{eq:GeneralQuadraticResponse}) can be applied to obtain the Arrhenius plot for the global rate. The corresponding plots for networks of size $20$, $100$ and $1.000$ are shown in Fig. \ref{fig:Histograms_and_Arrhenius}.
The convergence of the total energy distributions to a Gaussian is observed as the system size $N_v$ increases. For the parameters chosen here, the distribution appears already well described by a Gaussian for $N_v \sim 100$. The second derivative, i.e., twice the quadratic term as a function of $N_v$ is shown in Fig. \ref{fig:quadratic_vs_N}. One observes that as the system size increases, so does the quadratic term, in a statistical manner. This
proves numerically that this kind of networks generically lead to a quadratic Arrhenius plot in the limit
of large system size.

\begin{figure*}[p]
     \centering
    \includegraphics[width=1\textwidth]{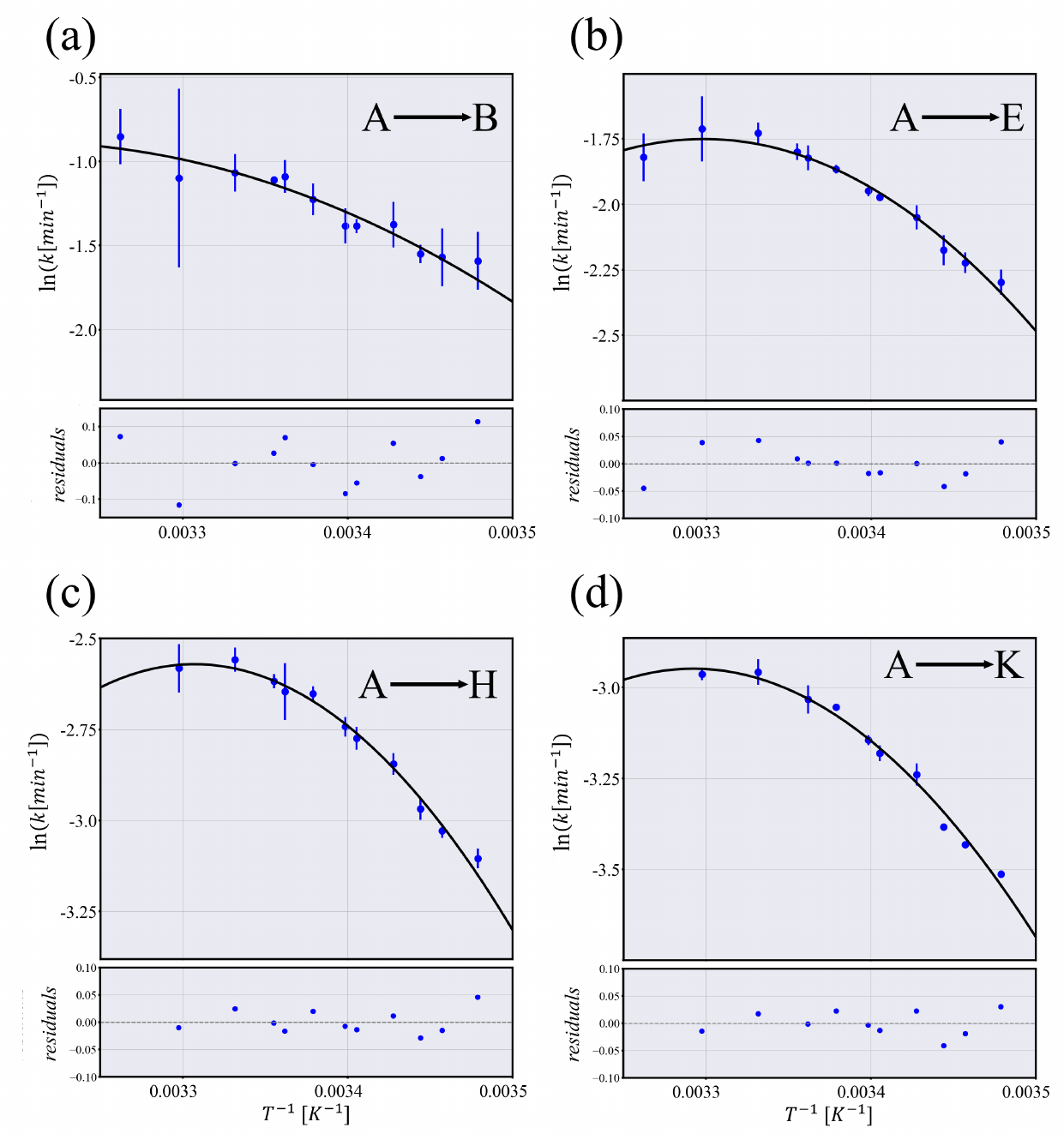}
     \caption{Arrhenius plot for the rates to the different stages in the development of \textit{Drosophila melanogaster} embryos, starting at the 14th nuclear division (A) adapted from \cite{DevelopmentalArrhenius} with quadratic fit to data. The residuals to the fit are shown below the main diagram.} The labeling of the stages follows the convention from the original authors: (a) rate from A to B (beginning of cellularization), $R^2=0.9587$ (b) rate from A to E (development of horizontal posterior midgut), $R^2=0.9909$ (c) rate from A to H (full germ band retraction), $R^2=0.9952$ (d) rate from A to K (first breath), $R^2=0.9868$.
     \label{fig:CrapseExamples}
\end{figure*}
\begin{figure*}[b]
    \centering
    \includegraphics[width=0.7\textwidth]{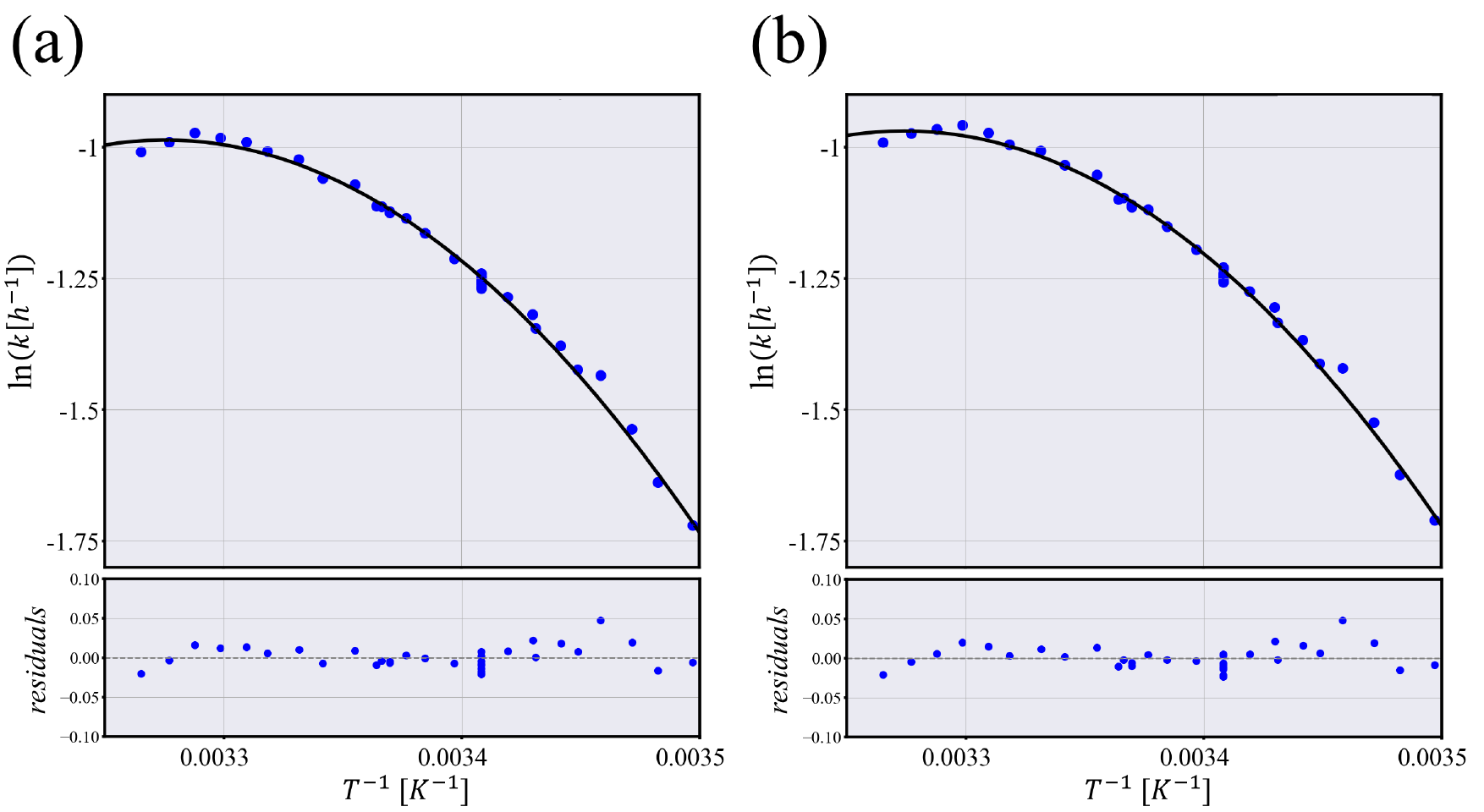}
    
    \caption{Logarithmic developmental rate to the pupal stage of \textit{Drosophila melanogaster} versus inverse temperature adapted from \cite{Bliss} with quadratic fit. The residuals to the fit are shown below the main diagram.} (a) For male specimen, $R^2=0.9950$ (b) for female specimen, $R^2 = 0.9950$.
    \label{fig:Bliss}
\end{figure*}

\begin{figure*}[p]
    \centering
    \includegraphics[width=1\textwidth]{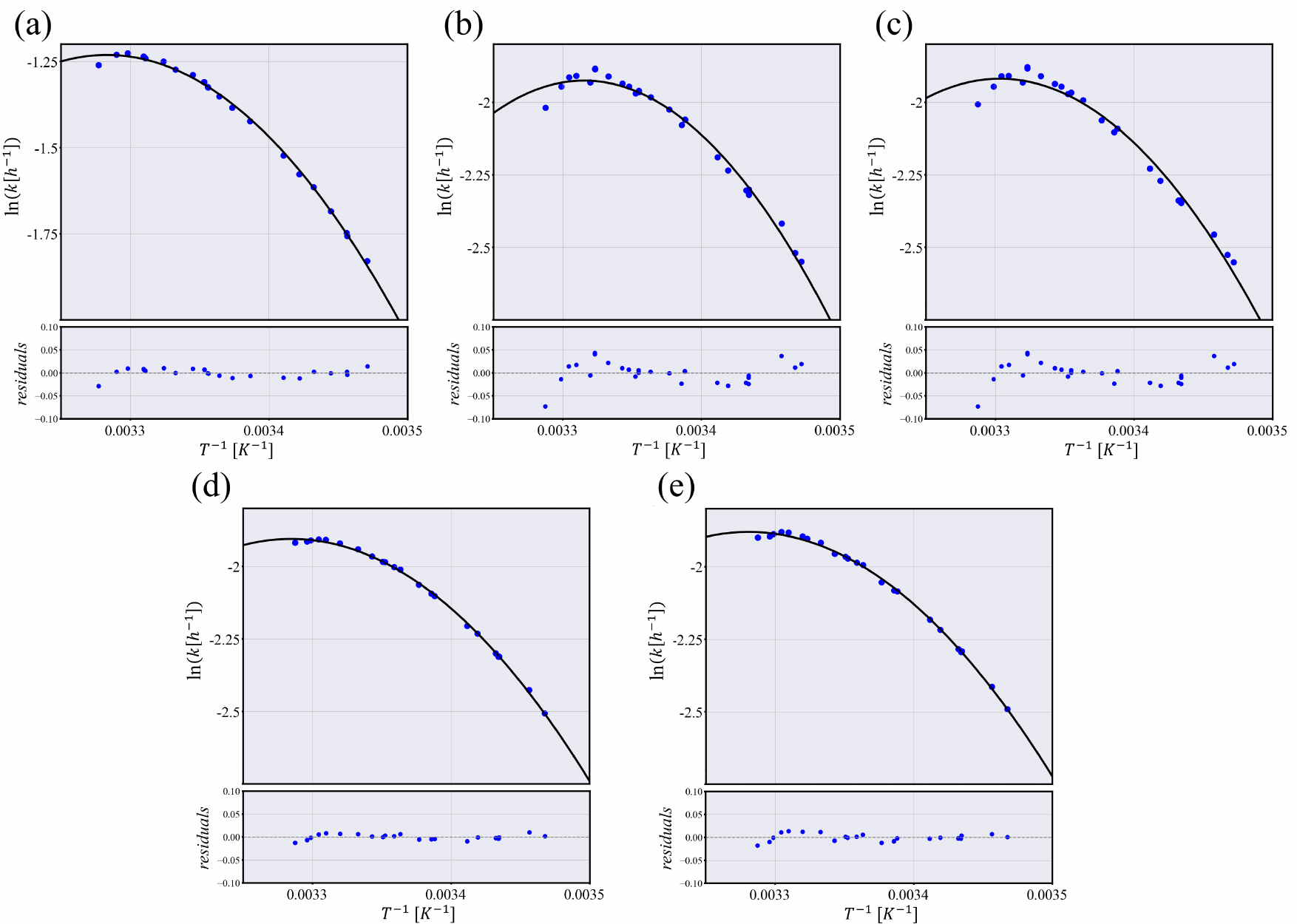}
    \caption{Logarithmic developmental rate for the embryonic, pupal and larval stage of \textit{Drosophila melanogaster} \cite{Powsner} versus inverse temperature with quadratic fit. The residuals to the fit are shown below the main diagram. (a) embyonic development, $R^2=0.9976$ (b) male larval development, $R^2 =0,9838$ (c) female larval development, $R^2=0.9808$ (d) male pupal development, $R^2=0.9990$ (e) female pupal development, $R^2=0.9981$.}
    \label{fig:Powsner}
\end{figure*}

\begin{figure}[ht]
     \centering
    \includegraphics[width=0.450\textwidth]{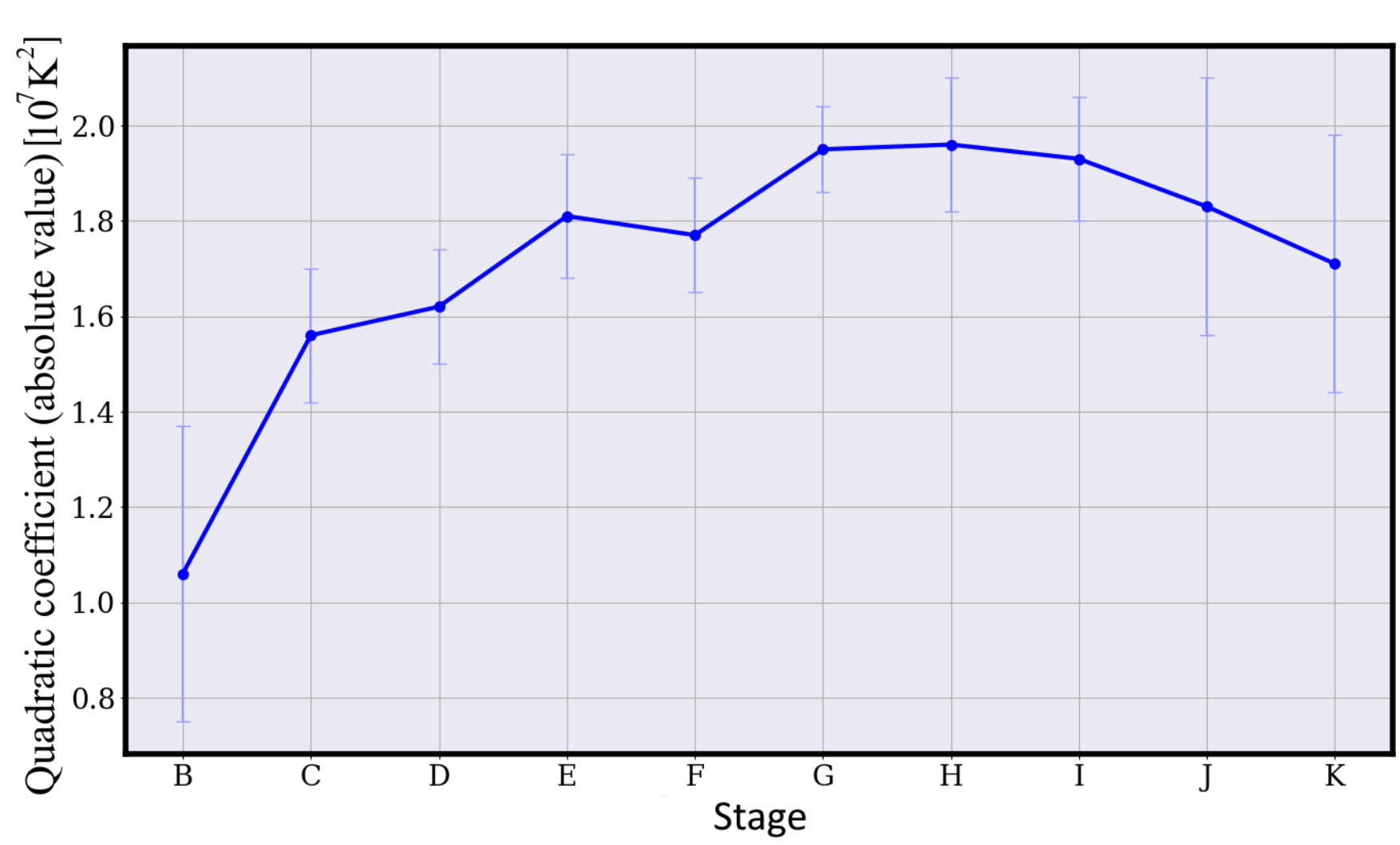}
     \caption{The absolute value of the quadratic coefficient from the quadratic fit to the data from \cite{DevelopmentalArrhenius} with the last cleavage as starting point to later stages in development.}
     \label{fig:Quadratic_Coefficient_Crapse}
\end{figure}

\subsection{Comparison to experimental data}

Our theoretical prediction is in striking agreement with experimental data for developmental systems. 
In a work published in 2021, Crapse et al. conducted a study on the development rates of \textit{Drosophila} embryos at different temperatures \cite{DevelopmentalArrhenius}. To that end, the embryonic phase of \textit{Drosophila}
was subdivided into 12 stages and the rates to reach each stage were measured under different temperature conditions. The 14th nuclear division in the zygote was chosen as the starting point (stage
A) and the first breath was chosen as the end point (stage K). The cumulative rates starting with the
14th nuclear division to 4 different stages (B,E,H,K in the notation of the original authors) are shown in Fig. \ref{fig:CrapseExamples} as an example, together with a quadratic fit performed in Python via the numpy.polyfit command. The data were taken from the supplementary material of the original publication, which tabulated the measured times for the entire sample and all stages. 
The fitting parameters are listed in Appendix \ref{app:FitParameters}. Note that these do not provide insights into the microscopic structure of the network without further assumptions, since the coefficients of the quadratic functions contain the differences of the first two moments along the trees and forests as well as terms containing $\frac{1}{k_BT_0}$.
Moreover, the residuals of the quadratic fit were computed to analyze whether there are any systematic deviations that hint towards a deviation from the quadratic dependence.

For the beginning of stage B (begin of cellularization), no robust statement can be made about the goodness of the quadratic fit. Since this is the first stage following the starting point, it may well be that the assumption of a sufficiently large underlying biochemical network is simply not valid. However, as more steps accumulate, such as in the rates to stages E, H, and K, the quadratic shape of the data points becomes more apparent. 
The residuals do not point to a systematic deviation, suggesting that higher order terms at best play a minor role.  
A quadratic fit of the data was also suggested by Crapse et al. \cite{DevelopmentalArrhenius}, although without providing an explanation on fundamental grounds, as done here.

Measurements of the temperature dependence of developmental rates of \textit{Drosophila} were already performed by Bliss \cite{Bliss} in 1926 for the prepupal development and by Powsner \cite{Powsner} in 1935 for the embryonic, pupal, and larval stages. Their results are shown in Fig. \ref{fig:Bliss} and \ref{fig:Powsner}, respectively, also with a quadratic fit to their data points. Both articles list their data in a table and show them in a diagram, albeit together with piecewise linear functions as fit instead of a global quadratic function.
One sees that in all cases, the quadratic fits predicted by our theory are in excellent agreement with
the experimental data. Also the residuals indicate only a minor contributions from terms of higher order, except for the larval development rates from the data by Powsner (Fig. \ref{fig:Powsner}b,c), where most likely some additional effect contributes.

Fig. \ref{fig:Quadratic_Coefficient_Crapse} shows the quadratic terms of the quadratic fit for the rates to the different stages defined by Crapse et al. \cite{DevelopmentalArrhenius}. Consistent with the simulation data shown in Fig. \ref{fig:quadratic_vs_N}, the quadratic term tends to increase as more steps accumulate. However, a slight decrease is observed for the rates to the final stages. This is most likely due to the lack of data points at the extreme ends of the temperature spectrum, as most embryos did not survive long enough under these conditions. Consequently, this also results in greater uncertainties for these values.
 
Note that the maximal rate, i.e.\ the peak in the diagrams, is at $T_{peak}\sim 30^\circ\text{C}$. Expecting the mean total activation energies to be higher for the trees than for the forests, and vice-versa for the variances, Eq. (\ref{eq:GeneralQuadraticResponse}) predicts the maximum to be attained at $\Delta\beta <0$, so $T_{peak}>T_0$, and one would in fact assume  $T_0\sim 20^\circ\text{C}$.

\subsection{Deviations from the generic quadratic response}
\label{sec:deviations}

Both our theory and computer simulations demonstrate that large networks show a generic Arrhenius plot with a quadratic form. 
\newnew{We now discuss deviations from this generic response
that might arise because of our mathematical assumptions.
We start with networks that have correlations
between their activation energies and show that
the generic response persists as long as such correlations
are sufficiently dilute in large networks.} We then discuss
small networks and demonstrate that they
easily can be designed to give non-linear Arrhenius plots.
We finally discuss the case of linear chains, which shows
that even large networks can deviate, because they 
contain only few spanning trees.

\newnew{\subsubsection{Correlations}
\label{app:IndependenceActivationEnergies}

Our theory assumes that activation energies on different edges are independent 
and that their summations over spanning trees and forests
satisfy the conditions of the central limit theorem. 
In real biochemical networks, the master equation description contains repeated reaction channels due to multiplicity, so the corresponding activation energies are in fact correlated. As an instructive example, consider an autorepressive production of species $X$:
\begin{align}
    \dot{X}= \frac{k}{1+(\frac{X}{C})^n},\label{eq:autorepressive_reaction}
\end{align}
where $k$ is the production rate in the absence of $X$, $C$ is a
reference concentration, and $n$ is the Hill coefficient. In a chemical master equation, the stochastic variable is the copy number $m$ of $X$ (for multiple species, it is the product space of all molecule counts). This yields 
a linear chain of reaction steps with rates $k_{m}=\frac{k}{1+(\frac{m}{c})^n}$, with $c$ being an appropriately rescaled constant (see Fig. \ref{fig:autorepressive_reaction_master_equation}). 
\begin{figure}[h]
     \centering
    \includegraphics[width=0.4\textwidth]{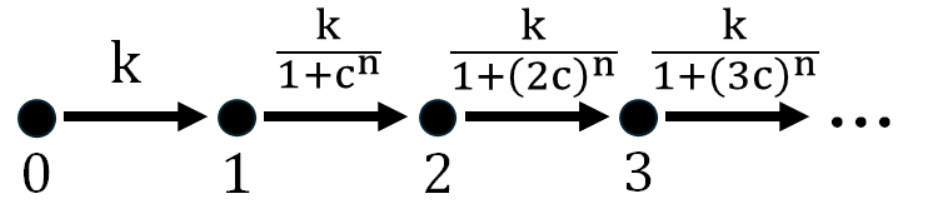}
     \caption{\newnew{Describing the autorepressive reaction in Eq. (\ref{eq:autorepressive_reaction}) as a master equation results in a linear chain of reactions where the rates all scale with the same reaction rate $k$.}}
     \label{fig:autorepressive_reaction_master_equation}
\end{figure}

Since the Arrhenius equation gives $k \sim e^{- \Delta\beta E_A}$, all these edges inherit the same activation energy $E_A$, because they are different instances of the same microscopic reaction. A more precise bookkeeping of total activation energy should therefore treat the energy of a spanning tree or forest as a weighted sum of distinct reaction types:
\begin{align}
    E_\mathcal{T}=\sum_{k_{ij}}n_{ij}E_{ij},\label{eq:sum_with_repeats}
\end{align}
where $n_{ij}\in\mathbb{N}$ counts how many times that same reaction appears in the tree or forest. Although the state space of such master equations can be infinite in principle, in real biological systems the number of copies is effectively bounded due to resource limitations and hence it can be safely assumed that the first two moments of $n_{ij}$ stay finite. Note that the multiplicities $n_{ij}$ can themselves be treated as independent random variables and consequently $n_{ij}E_{ij}$ are independent random variables again. As long as there are still sufficiently many distinct, effectively independent reactions, Eq. (\ref{eq:sum_with_repeats}) is still the sum of many independent random variables and the central limit theorem retains its validity.

This argument also applies to the possibility that the same network motif
appears several times in the network, as known for example from 
the feedforward chain appearing in the developmental network of \textit{Caenorhabditis elegans} \cite{ewe2022feedforward}. Again this case could be remedied by introducing multiplicities $n_{ij}$.
Only if the network was highly correlated throughout, 
would our generic prediction break down.}

\subsubsection{Small networks}

\begin{figure*}[bht]
     \centering
    \includegraphics[width=0.8\textwidth]{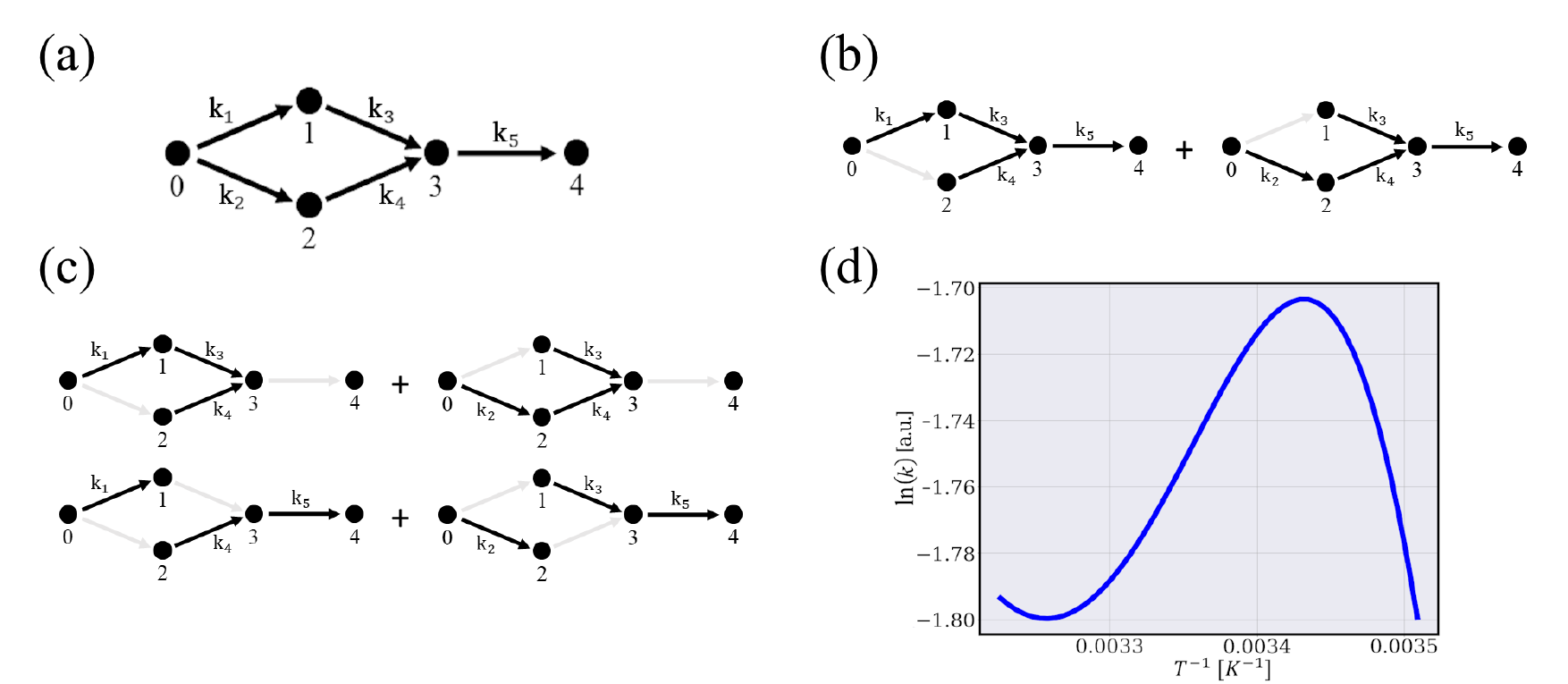}
     \caption{Example network on five vertices with a strong deviation from an Arrhenius-like response. (a) Network (b) Spanning trees (c) spanning forests (d) Arrhenius plot for the ratio $(k_1:k_2:k_3:k_4:k_5) =(2:2:5:0.1:2)$ at $T_0 = 300K$ and $E_1= 20 \frac{\text{kJ}}{\text{mol}}$, $E_2= 100 \frac{\text{kJ}}{\text{mol}}$, $E_3= 30 \frac{\text{kJ}}{\text{mol}}$, $E_4= 20 \frac{\text{kJ}}{\text{mol}}$ and $E_5= 80 \frac{\text{kJ}}{\text{mol}}$.}
     \label{fig:Arrhenius_strong_deviation}
\end{figure*}

As an example for a small network, consider the $N_v = 5$ network
depicted in Fig. \ref{fig:Arrhenius_strong_deviation}a. 
This network has two spanning trees rooted at vertex $4$ (see Fig. \ref{fig:Arrhenius_strong_deviation}b):
\begin{align}
    Z_{\mathcal{T}}=(k_1+k_2)k_3k_4k_5,
\end{align}
and four two-tree spanning forests, where one tree is rooted at vertex $4$ and the other contains vertex $0$ (see Fig. \ref{fig:Arrhenius_strong_deviation}c):
\begin{align}
    Z_{\mathcal{F}}=k_1k_4(k_3+k_5)+k_2k_3(k_4+k_5).
\end{align}

Thus, the MFPT from $0$ to $4$ reads:
\begin{align}
    \langle\tau\rangle =\frac{Z_{\mathcal{F}}}{Z_{\mathcal{T}}}=\frac{1}{k_1+k_2}\Big(\frac{k_1}{k_3}+\frac{k_2}{k_4}\Big)+\frac{1}{k_5},
\end{align}
which can give rise to the dependence in the Arrhenius plot shown in Fig. \ref{fig:Arrhenius_strong_deviation}d for realistic values for the rate constants and activation energies.

The resulting temperature response no longer resembles the linear slope of the Arrhenius equation; instead, it is strongly curved, reaching a minimum around $T=35^\circ C$ and a maximum around $T=18^\circ C$. This implies that the global rate would actually decrease with increasing temperature within the temperature range typically relevant to biological systems. This example illustrates that the temperature dependence in medium-sized networks depends on microscopic details and may generally be more complex.

\begin{figure}[h]
     \centering
    \includegraphics[width=0.3\textwidth]{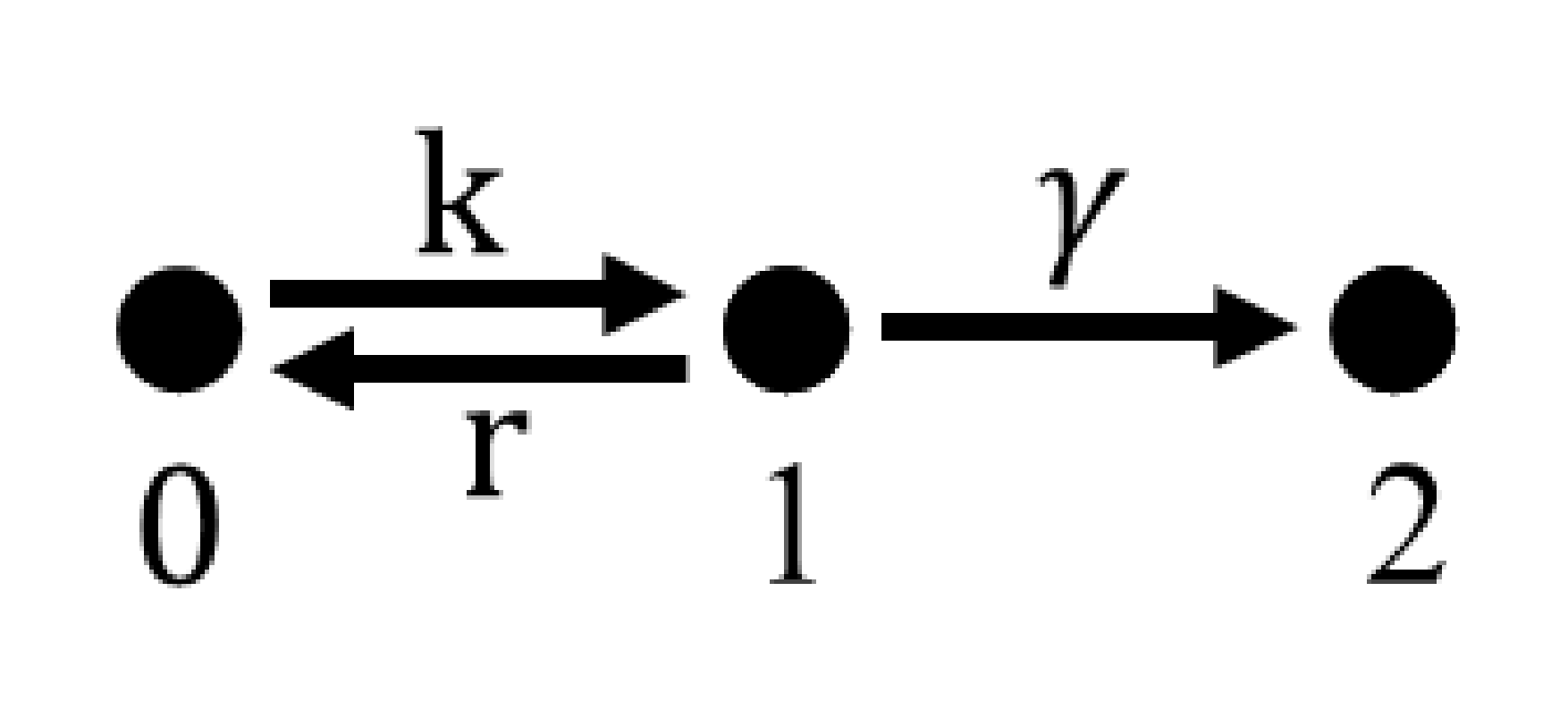}
     \caption{A minimal example to show that a network with a rate-limiting step does not necessarily yield a global Arrhenius equation dominated by this step.}
     \label{fig:minimal_example}
\end{figure}

An interesting case are networks with a single rate-limiting step, i.e.\ a single reaction that proceeds significantly slower than the rest of the reaction network. In this case, one might naively expect that the Arrhenius equation for this reaction dominates and that the global response becomes linear again. However, this is not necessarily the case. Consider the minimal example shown in Fig. \ref{fig:minimal_example}.
Observe that: 
\begin{equation}
    \langle\tau\rangle=\frac{Z_\mathcal{F}}{Z_\mathcal{T}}=\frac{k+r+\gamma}{k\gamma}=(1+\frac{r}{k})\frac{1}{\gamma}+\frac{1}{k}.
\end{equation}
If the step from $1$ to $2$ is rate-limiting, i.e., $\gamma \ll k,r$, then:
\begin{equation}
    \langle\tau\rangle\approx(1+\frac{r}{k})\frac{1}{\gamma},
\end{equation}
not just $\frac{1}{\gamma}$ as one could naively expect. So, the ratio $\frac{r}{k}\sim e^{\beta\Delta E}$ also contributes even if $\gamma$ is many orders of magnitude below $k$ and $r$. 
The reason for this is that temperature also impacts the equilibrium distribution. If $\gamma$ is the rate-limiting step, one can assume that the states $0$ and $1$ are effectively in their equilibrium configuration, meaning that the flow from $0$ to $1$ must be equal to the flow from $1$ to $0$. Denoting the probability to be in state $1$ as $p$, this means:
\begin{align}
    k(1-p)=rp\Rightarrow p=\frac{k}{k+r}.
\end{align}
The crucial point is that the system can only jump to $2$ via $1$, and the rate must therefore be multiplied by the probability $p$ to be in $1$. Hence, one gets:
\begin{align}
    \langle \tau \rangle = (p\gamma)^{-1}=(1+\frac{r}{k})\frac{1}{\gamma},
\end{align}
which is the same result as before. One can therefore conclude that separation of time scales is not sufficient to obtain a pure Arrhenius dependence.

\subsubsection{Linear chain}

\begin{figure}[h]
\includegraphics[width=0.45\textwidth]{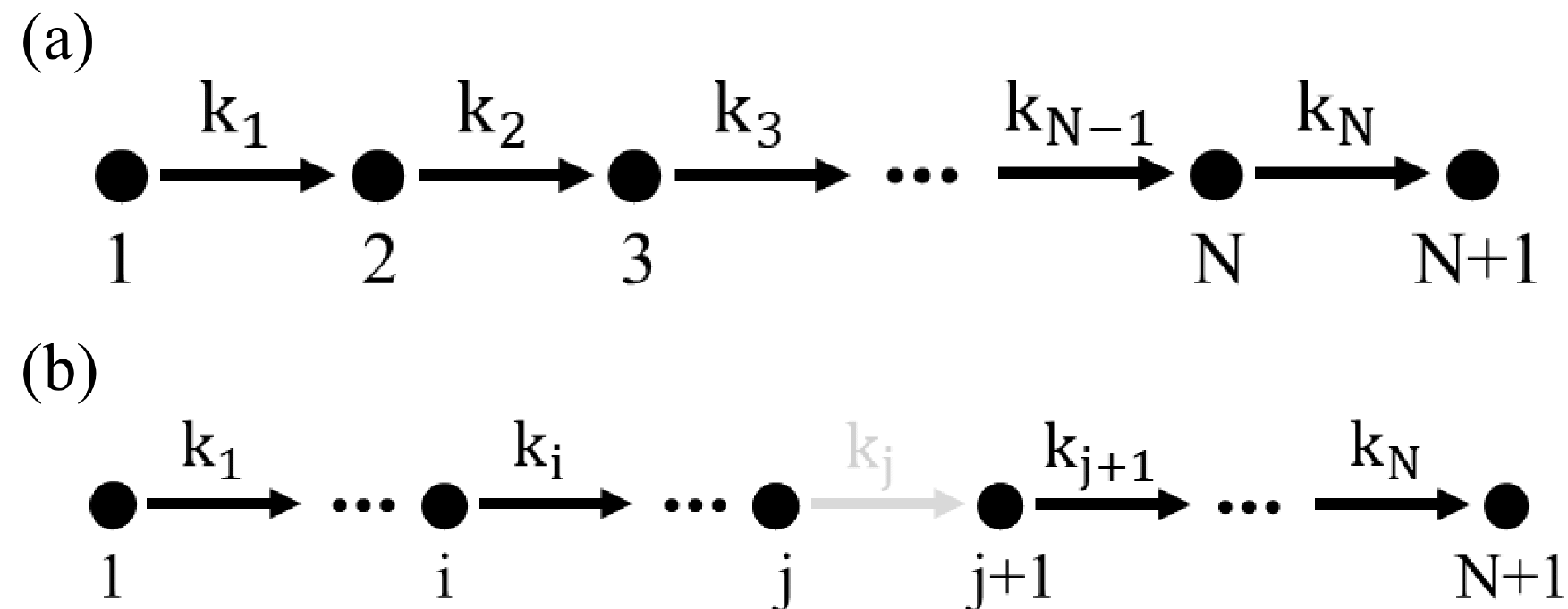}
\caption{\label{fig:LinChain} (a) A linear chain of $N$ reactions with reaction rates $k_{i,j}=k_i\delta_{i+1,j}$ (b) Spanning forests of the linear chain. For $i\leq j$, the only corresponding spanning forest is the one without the $k_j$-edge. For $i>j$, there is no such tree.}
\end{figure}

The generic quadratic behaviour predicted here can
be violated not only for small networks, but also 
for large networks which contain only few spanning
trees. The simplest possible example is a linear chain of irreversible reactions, i.e., $k_{i,j}=k_i\delta_{i+1,j}$, as illustrated in Fig. \ref{fig:LinChain}.

Then the corresponding graph is already a tree and thus its own unique spanning tree: 
\begin{equation}
    \sum_{\mathcal{T}_{[N+1]}}w(\mathcal{T})=\prod _{i=1}^{N}k_i.
\end{equation}
Also, for the spanning forests it is easy to see that there is either exactly one or no combination:
\begin{equation}
    \sum _{\mathcal{F}_{[j,N+1]}^{i\rightarrow j}}w(\mathcal{F})=\frac{1}{k_j}\prod _{i=1}^{N}k_i\delta_{i\leq j}.
\end{equation}
Then,
\begin{equation}
    (-K^{T})^{-1}=\left( \begin{matrix}
k_1^{-1} & k_2^{-1}  & \dots & k_{N}^{-1} \\
 0 & k_2^{-1}  & \dots & k_N^{-1}\\
 \vdots &\vdots& \ddots & \vdots \\
 0 & 0  & 0 &k_N^{-1}
\end{matrix} \right).
\end{equation}
One finds for the first two moments:
\begin{align}
    \langle\tau_i\rangle &=\sum _{j=i}^N\frac{1}{k_j},\label{eq:MFPT_Linear_Chain}\\\langle\tau_i ^2\rangle &=2\sum _{j=1}^N\frac{1}{k_j}\sum _{l=j}^N\frac{1}{k_l}=\Big (\sum _{j=1}^N\frac{1}{k_j}\Big )^{2}+\sum _{j=1}^N\frac{1}{k_j^2},
\end{align}
which yields the standard deviation:
\begin{equation}
    \sigma_{\tau_1}=\Big (\sum _{j=1}^N\frac{1}{k_j^2}\Big)^{\frac{1}{2}}
\end{equation}
and the coefficient of variation (CV) reads:
\begin{equation}
    CV_{\tau_1}=\frac{\sigma_{\tau_1}}{\langle\tau_1\rangle}=\Bigg(\frac{\sum _{j=1}^N\frac{1}{k_j^2}}{(\sum _{j=1}^N\frac{1}{k_j})^2}\Bigg)^{\frac{1}{2}},\label{eq:CVforLinChain}
\end{equation}
meaning that $CV_{\tau_1}\to 0$ for $N\to\infty$ if most of the $k_i$ are of comparable sizes\footnote{One can find mathematical counterexamples otherwise, e.g., $k_i=i^2k$, which converges to $CV=\frac{2}{\sqrt{10}}$}, so the relative variation vanishes in the limit of large chains.

If $k_i=k$ $\forall i$, one finds in general:
\begin{equation}
    \langle\tau ^n\rangle_i=\frac{1}{k^n}\frac{(N+n-i)!}{(N-i)!},
\end{equation}
which are precisely the moments for an $N-i+1$-dimensional Erlang distribution ($\Gamma$-distribution with non-negative integer as shape parameter) with $\lambda =\frac{1}{k}$:
\begin{equation}
    f^{\tau_i}(t)=\frac{k^{N-i+1}}{(N-i+1)!}t^{N-i}e^{-kt},
\end{equation}
as it should be since this is precisely the distribution that arises from summing up $N-i$ iid exponentially distributed random variables.

Note that the coefficient of variation (CV) reads:
\begin{equation}
    CV_{\tau_1}=\frac{\sigma_{\tau_1}}{\langle\tau_1\rangle}=\frac{1}{\sqrt{N}}\overset{N\to\infty}{\to} 0,
\end{equation}
so the relative deviation from the mean becomes increasingly small as the system size increases.

Moreover, the mean first-passage time for the whole process, $\langle\tau\rangle =\langle\tau_0\rangle$, given in Eq. (\ref{eq:MFPT_Linear_Chain}) may be written in the limit of large $N$ as:
\begin{align}
    \langle\tau\rangle &=N\langle k_j^{-1}\rangle=N\langle {k^0_{j}}^{-1}e^{\Delta\beta E_{j}}\rangle,
\end{align}
writing $k_j$ as in Eq. (\ref{eq:AltArrhenius}). Assuming $k^0_j$ and $E_{j}$ to be independent, one finds:
\begin{align}
    \ln{\langle\tau\rangle} &=\ln{\langle e^{\Delta\beta E_{j}}\rangle}+\ln{N}+\langle {k^0_{j}}^{-1}\rangle \\ &= \ln{\langle e^{\Delta\beta E_{j}}\rangle}+const.\label{eq:LinChainStrongLaw}
\end{align}
Compare this to Eq. (\ref{eq:GlobalArrheniusGeneralFormula}): for the linear chain, there was only a single spanning tree, so $\kappa _1^{E_{\mathcal{T}}}=E_{\mathcal{T}}$ and $\kappa _n^{E_{\mathcal{T}}}=0$ for $n\geq 0$. The total activation energies of the forests $\mathcal{F}_i$ for $i=1,...,N$ read:
\begin{equation}
    E_{\mathcal{F}_i} = \sum\limits_{\substack{j=1 \\ j\neq i}}^NE_j=E_{\mathcal{T}} -E_i.
\end{equation}
One then obtains for the cumulant-generating function of $E_{\mathcal{F}}$:
\begin{align}
    \sum \frac{\Delta \beta^n}{n!}\kappa^{E_{\mathcal{F}}}_n&=\ln{\langle e^{\Delta \beta E_{\mathcal{F}}}\rangle}\\ &=\ln{e^{\Delta \beta E_{\mathcal{T}}}+\ln{\langle e^{- \Delta \beta E_i}\rangle}}\\ &=\Delta \beta E_\mathcal{T}+\sum_{n=1}^\infty \frac{\Delta \beta ^n}{n!}\kappa^{E_i}_n.
\end{align}
Inserting this into Eq. (\ref{eq:GlobalArrheniusGeneralFormula}) yields:
\begin{align}
    \ln{\langle\tau\rangle} &= \Delta \beta E_\mathcal{T}+\sum_{n=1}^\infty\frac{\Delta \beta ^n}{n!}\kappa^{E_i}_n-\Delta \beta E_\mathcal{T}+const.\\ &= \ln{\langle e^{\Delta\beta E_{j}}\rangle}+const.,
\end{align}
so indeed the same result as Eq. (\ref{eq:LinChainStrongLaw}).

The total activation energy of the forests only differs from the total activation energy of the trees by a single $E_i$ and only the cumulant generating function of $E_j$ but not a sum over many activation energies remains. Therefore, there is no longer a statistical limit $N\to \infty$ for the distribution and the expression will depend on the distribution of $E_i$. If $E_i$ is itself normal distributed, the quadratic dependence in Eq. (\ref{eq:GeneralQuadraticResponse}) is recovered. But choosing a uniform distribution $E_i\sim U([E_{\text{min}},E_{\text{max}}])$, one obtains:
\begin{align}
    \langle e^{\Delta\beta E_{j}}\rangle & =\frac{1}{\Delta E}\int_{E_{\text{min}}}^{E_{\text{max}}}\text{d}E_j e^{\Delta\beta E_{j}} \\ & =\frac{1}{\Delta E\Delta\beta}( e^{\Delta\beta E_{\text{max}}}-e^{\Delta\beta E_{\text{min}}}),
\end{align}
with $\Delta E:= E_{\text{max}}-E_{\text{min}}$.
Then Eq. (\ref{eq:LinChainStrongLaw}) reads:
\begin{align}
    \ln{\langle\tau\rangle}&= -\ln{\Delta E\Delta\beta} + \Delta\beta E_{\text{min}} +\ln{(e^{\Delta\beta \Delta E}-1)}\nonumber\\&=\Delta\beta\frac{E_\text{max}+E_\text{min}}{2} +\sum _{n=2}^\infty\frac{\Delta E ^n}{n!}\frac{B_n}{n}\Delta \beta ^n ,
\end{align}
where $B_{n}$ is the $n$-th Bernoulli number. 
This corresponds to the simulations performed by Crapse et al. \cite{DevelopmentalArrhenius}, involving a chain of $1.000$ reactions with activation energies drawn from a uniform distribution. The authors found that the results of their simulation could not reproduce the quadratic shape that they found describing their experimental data and therefore discarded the possibility that it was a result of the complexity of the network. In light of the analytical theory developed here, that choice seems unfortunate because the linear chain is too simplistic to allow for the large system limit. Additionally, if they had chosen $E_i\sim \mathcal{N}(\langle E\rangle , \sigma ^2)$, it would have resulted in a quadratic shape.

\section{Discussion}

Here we have shown that the generic temperature response of large biochemical networks
is quadratic in the Arrhenius plot if the network
has an imbalance of activation energies in forward
versus backward directions. This seems reasonable for developmental processes and is supported by evidence from \textit{in vitro} studies with enzymes from \textit{Xenopus} showing a substantial imbalance in the activation energies of the cyclin synthesis and degradation rates \cite{rombouts2024mechanistic}.

In order to make general
statements on such networks, we employed a statistical description which does not require
knowledge of all activation energies. Instead, we only make assumptions about their distribution, which is a reasonable description for a large system size $N$, where not all degrees of freedom can be known. Our discussion was based on a general expression of the MFPT for arbitrary networks, Eq. (\ref{eq:MFPTFormula}), which follows from graph theory \cite{FPTGunawardena} and has been derived here
in a pedagogical manner. We then rewrote this
formula in the language of statistical physics.
The coefficients appearing in the Taylor-series of the logarithmic MFPT can be interpreted in terms of the cumulants of the distribution of the sums along the spanning trees and two-tree forests. If this sum can be sufficiently well approximated by a Gaussian distribution, this yields a quadratic dependence in the Arrhenius plot, as described by Eq. (\ref{eq:GeneralQuadraticResponse}). This is our main result, namely the quadratic
response as the natural description for complex biochemical networks. This result need not apply for a network
which has similar distributions for the activation
energies for forward and backward directions, but this
is unlikely for developmental systems.

Being the result of evolution, the assumption that the activation energies can be described as independent from each other needs to be treated with care. \newnew{We first note that correlations arising from copies numbers in chemical
master equations are not a fundamental limitation
of our theory, because this effect can be described
by multiplicities in the relevant sums. We also note that
duplication of the same motif would not matter for the same reason.
However, it could well be that two network motifs have been
duplicated and then drifted apart during evolution.
To date, no experimental evidence
seems to be available to determine how strong
such correlations might be in real developmental networks.}
However, since there are many different reactions involved in a process such as the development of an embryo, some of
which are  purely biochemical (like the phosphorylations
occuring in cell cycle control), some of which are more biophysical
in nature (like  the growth of microtubules), it seems
reasonable to assume that most activation energies are
indeed independent of each other, especially when contained
in different modules of the overall network. 
Furthermore, generalizations of the central limit theorem exist that allow for a weak degree of dependence, e.g. correlations
decaying with distance in the network (see for example \cite{bradley2007introduction}). In the future, 
it might be interesting to experimentally determine
all activation energies in a network of interest and
then to predict the resulting temperature dependence
using the formalism presented here.

The theory was validated numerically by drawing spanning trees and two-tree spanning forests from a random graph on $N_v = 1.000$ vertices.
Individual activation energies were drawn from homogeneous distributions, also 
in order to avoid using a Gaussian distribution from the start. 
We then computed the histograms of their total activation energies, which approach a normal distribution as expected from the central limit theorem (Fig. \ref{fig:Histograms_and_Arrhenius}). The global rate constant in an Arrhenius plot is then of quadratic shape, and this quadratic dependence becomes more pronounced for larger $N_v$, as predicted
by the theory (Fig. \ref{fig:Histograms_and_Arrhenius}).

Crapse et al. \cite{DevelopmentalArrhenius} suggested that the observed deviation from a pure Arrhenius-equation was due to the biochemical reactions not following an ideal Arrhenius-behavior, rather than a result of the complexity of the system. They reached that conclusion based on the simulation of a linear chain of $1.000$ reactions described by an Arrhenius-equation, which failed to reproduce the observed quadratic response in the Arrhenius plot. They also conducted an experiment measuring the conversion of NAD+ to NADH via GAPDH catalysis and found that this reaction does not follow a pure Arrhenius-like dependence. 
However, closer inspection of their simulations shows that the activation energies were chosen uniformly in an energy range from $20\frac{\text{kJ}}{\text{mol}}-100\frac{\text{kJ}}{\text{mol}}$ and the prefactors independently at $T_0=295\text{K}$. Due to the absence of backward rates and branching in this model, this leads to a different series (see section \ref{sec:deviations}), for which the uniform distribution yields infinitely many non-vanishing terms. If the activation energies were chosen from a normal distribution, this would yield a quadratic dependence also on this level. In contrast, our results are more general and require fewer assumptions. 
Moreover, the measurement of the temperature dependence of the rate of the NAD+ to NADH reaction seems to be well described by a linear fit in the Arrhenius-plot up to at least $35^\circ\text{C}$. Indeed, for higher temperatures, the denaturation of the involved enzymes significantly decreases the reaction rate. The measurements of the development rates only reached about $33^\circ\text{C}$, though. Therefore, this can most likely not explain the quadratic temperature dependence observed over the entire range.

The theory developed so far is of course no longer valid above the melting temperature of the proteins. Denaturation substantially slows the reaction rates, and this is not described by the Arrhenius equation \cite{DevelopmentalArrheniusRanges,ProteinDenat}. This could explain the curvature of the Arrhenius plots seen
at small values of the inverse temperature,
in particular for temperatures of about $30^\circ\text{C}$ and above, where we observe that the high temperature data points (low inverse temperature) in Fig. \ref{fig:CrapseExamples}-\ref{fig:Powsner} all lie below the quadratic curve.
In the measurement by Powsner on the pupal development rates (Fig. \ref{fig:Powsner}b), the decrease is most notable, and it is not unlikely that this indicates that the denaturation threshold for a protein that is important for the developmental progression is surpassed at these temperatures. On the other hand, however, this observation also suggests that the larger part of the curve should not be affected by protein denaturation, and that our theory indeed captures the generic behavior of this system. Future work could extend our theory to include the effects of protein denaturation, for example, by introducing it as an additional state in the network.

It should be mentioned that Powsner already noticed in his work that the combination of a few reactions, each described by an Arrhenius equation, does not necessarily yield a global Arrhenius equation and hypothesized that the observed temperature response may be a consequence of the complexity of the underlying biochemical network. However, he did not give a mathematical description as provided here. 

The observation that the development rates can be well described by a quadratic fit in the Arrhenius plot does not seem to be limited to fly and frog as discussed above. Rombouts et al. used it to fit the timing of the early cell cycles of \textit{Caenorhabditis elegans}, \textit{Caenorhabditis briggsae}, \textit{Danio rerio},
\textit{Xenopus tropicalis} and \textit{Xenopus laevis} \cite{rombouts2024mechanistic}, and Aquilanti et al. found that a quadratic dependence in the Arrhenius plot is a good description for data on the respiration rate of \textit{Camellia Japonica} leaves \cite{aquilanti2010temperature}.

Our result is a statement about large biochemical networks. Due to the exponential nature of the Arrhenius equation, it is reasonable to assume that this effect dominates more subtle temperature dependencies,
like the dependence of the diffusion constant on temperature.
According  to the Einstein relation, the diffusion constant $D$ is linear in $T$ and a change from $T = 300K$ to $T=310K$ increases $D$ by a mere 3 \%, while chemical rates often double or triple under a change of $\Delta T =10K$ \cite{elias2014universality}. Therefore, it seems justified to compare the theory to data on developmental processes, without claiming to capture every nuance of the effect of temperature on these complex systems. If needed, the effect of temperature on the diffusion
constant can be included by decomposing binding rates into
diffusion and reaction parts \cite{gomez2023grand}.

The discussion in this work was limited to the mean of the first-passage time, although the
formalism in principle covers all higher moments, too. 
The focus on the first moment seems justified by the observation that the coefficient of variation (standard deviation divided by mean) for the development times of \textit{Drosophila} embryos is in the order of a few percent across different temperatures \cite{LowCV}. This results most likely from a strong forward bias for the rates of a developmental network with many steps like cell cycle checkpoints and cell division being practically irreversible. It is well known that this tends to yield sharp FPT distributions around the mean in the limit of large networks \cite{SimplicityComplexNW} (compare also the example in section \ref{sec:deviations}). 

The framework developed here was
motivated by describing the temperature dependence of complex networks where not all
degrees of freedom are known. In this work, we focused on development, but as mentioned in the introduction,
there exist other complex biological systems with interesting temperature dependence, in particular
in the contexts of fever and climate change. 
Our results are relatively general statements about
the mean first-passage time of master equations and are therefore not \textit{per se} limited to
biochemical networks and temperature effects. The algorithm used to obtain the distributions of total activation energies along trees and two-tree forests also provides a method for calculating the MFPT for large networks. It may be worthwhile to explore how this compares to obtaining the MFPT from simulations of stochastic dynamics \cite{gillespie2007stochastic}, and whether it can also be applied to study other network properties related to graph theory. 

The idea to describe large reaction networks from a coarse-grained perspective with appropriate approximations has been discussed
elsewhere \cite{LargeNWBook}, but neither the mean first-passage time nor the possibility of describing the rate constants on a statistical basis as proposed here seem to be explored so
far. First-passage time problems are ubiquitous in biology and biochemistry \cite{MFPTsBiology,ham2024stochastic}
and it is very interesting how far one can get without knowing all the details of the system under consideration.

\begin{acknowledgments}

JBV thanks the German Academic Scholarship Foundation (Studienstiftung des Deutschen Volkes) for support. 
We also acknowledge support by the
Max Planck School Matter to Life funded by the German Federal Ministry of Education and Research (BMBF) 
in collaboration with the Max Planck Society. 

\end{acknowledgments}

\appendix




\section{Obtaining the MFPT by counting}
\label{app:MFPTCountingExample}
To illustrate the application of the graph-theoretical approach used here, 
we derive the MFPTs for two examples which also can be obtained by applying Laplace-transforms \cite{SimplicityComplexNW}.

\subsubsection*{One-step master equation}
\begin{figure}[h]
    \centering
    \includegraphics[width=0.45\textwidth]{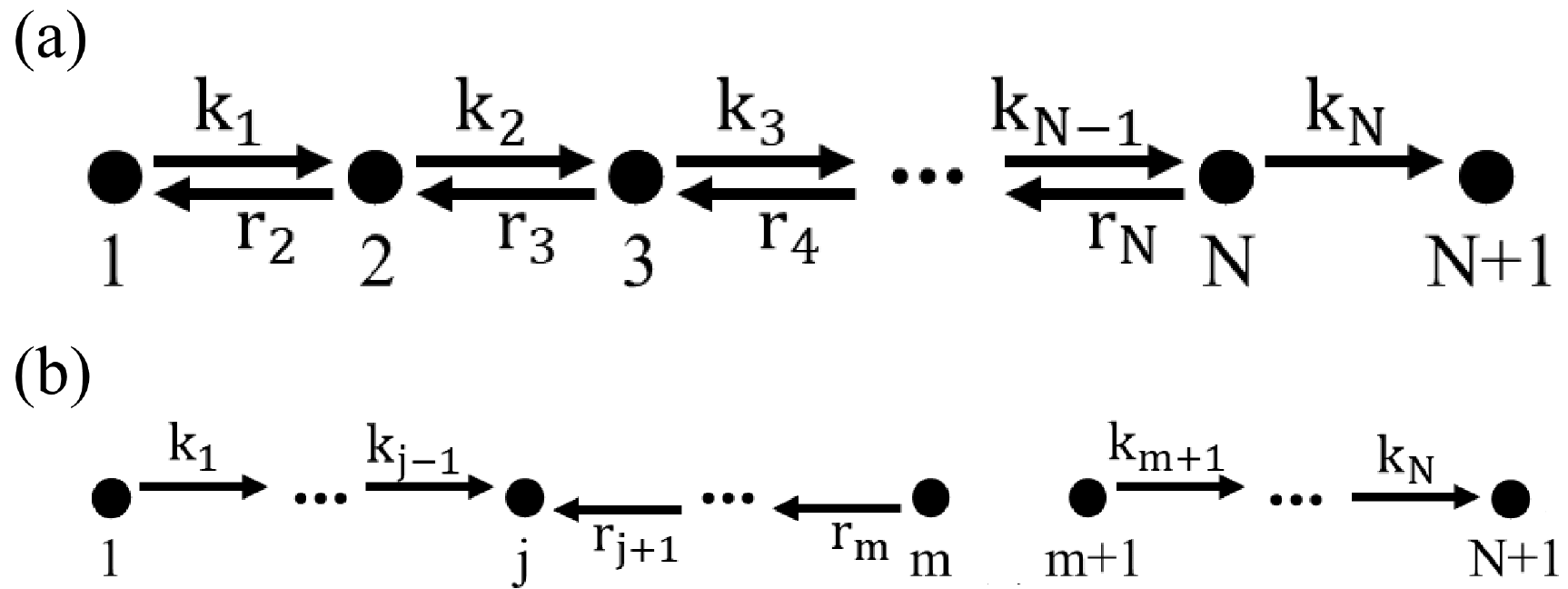}
    \caption{(a) A one-step master equation from state $1$ to state $N+1$, the latter one being absorbing. \newline(b) Counting the two-tree spanning forests rooted at $N+1$ $\mathcal{F}_{[j,N+1]}^{1\rightarrow j}$ for this network.}
    \label{fig:OneStep}
\end{figure}

Consider the one-step master equation with forward rates $k_i$ and backward rates $r_i$ as depicted in Fig. \ref{fig:OneStep}. 

Note that there is only one spanning tree with all flow directed towards $N_v$, namely the one consisting only of the forward rates, meaning that:
\begin{equation}
   \sum_{\mathcal{T}_{[N+1]}}w(\mathcal{T})=\prod _{i=1}^{N}k_i.
\end{equation}
The $\mathcal{F}_{[j,N+1]}^{1\rightarrow j}$ are the chains which start off towards the right and change their direction at $j$ and possibly also at a later vertex $m$. This is shown in Fig. \ref{fig:OneStep}b.
Therefore, one gets:
\begin{equation}
\sum_{j=1}^N\sum _{\mathcal{F}_{[j,N+1]}^{1\rightarrow j}}w(\mathcal{F})=\sum_{m=1}^N\sum_{j=1}^m\prod_{\alpha=1}^{j-1}k_\alpha\prod_{\beta=l+1}^mr_\beta\prod_{\gamma=m+1}^N k_\gamma.
\end{equation}
By Eq. (\ref{eq:MFPTFormula}), the MFPT is thus given by:
\begin{equation}
    \langle\tau\rangle =\sum_{i=1}^{N}\sum_{l=1}^i\frac{1}{k_l}\prod_{m=l+1}^i\frac{r_m}{k_m}\label{eq:MFPTOneStep}.
\end{equation}

\subsubsection*{Simple Kinetic Proofreading}

\begin{figure}[h]
    \centering
    \includegraphics[width=0.45\textwidth]{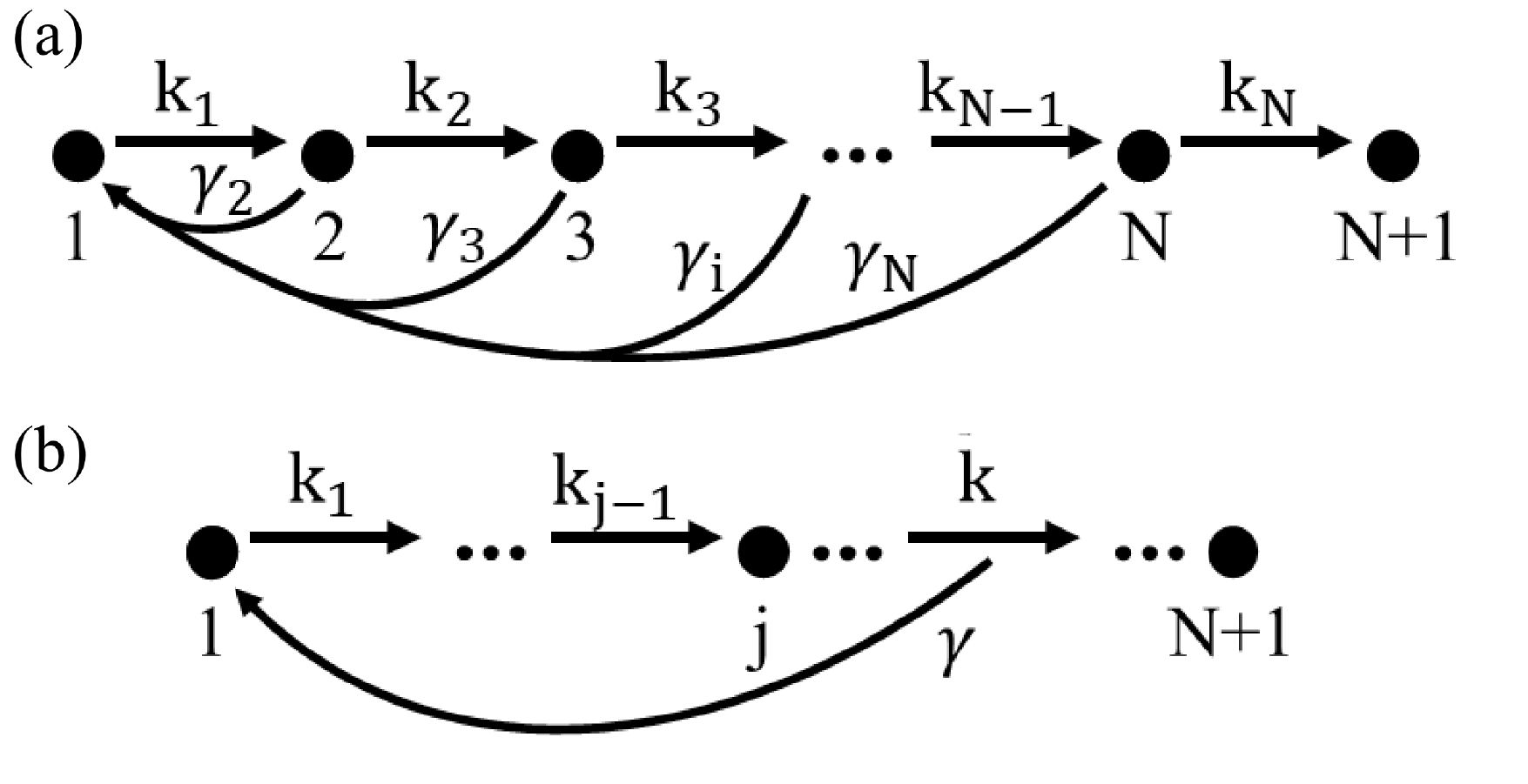}
    \caption{(a) A simple KPR network on $N+1$ states consisting of a chain of forward rates $k_i$ and reset rates $\gamma_i$. (b) Counting the $\mathcal{F}_{[j,N+1]}^{1\rightarrow j}$ for this network.}
    \label{fig:KPRSimple}
\end{figure}

One can also derive the MFPT for a simple kinetic proofreading (KPR) scheme shown in Fig. \ref{fig:KPRSimple} by correct counting.
Again, there is just one spanning tree, namely the one containing all $k_i$. So one has again:
\begin{equation}
   \sum_{\mathcal{T}_{[N+1]}}w(\mathcal{T})=\prod _{i=1}^{N}k_i.
\end{equation}
The $\mathcal{F}_{[j,N+1]}^{1\rightarrow j}$ are the graphs with forward rates up to some vertex $j$ and either $k_i$ or $\gamma_i$ for any $i>j$. This is illustrated in Fig. \ref{fig:KPRSimple}b.
Therefore, one gets:
\begin{equation}
    \sum_{j=1}^N\sum _{\mathcal{F}_{[j,N+1]}^{1\rightarrow j}}w(\mathcal{F})=\sum _{j=1}^{N}\prod _{i=1}^{j-1}k_i\prod_{m=j+1}^{N}(k_m+\gamma _m)
\end{equation}
and obtains for the MFPT:
\begin{equation}
    \langle\tau\rangle =\sum _{j=1}^{N}\frac{1}{k_j}\prod_{m=j+1}^{N}(1+\frac{\gamma _m}{k_m}).
\end{equation}
For the case that $k_i=k$ and $\gamma _i =\gamma$, this expression simplifies to:
\begin{align}
    \langle\tau\rangle =\sum _{j=1}^{N}\frac{1}{k}(1+\frac{\gamma }{k})^{N-j} =\frac{(1+\frac{\gamma }{k})^{N}-1}{\gamma}.
\end{align}
 
\section{Fitting parameters for the figures in the main text}\label{app:FitParameters}

The parameters for the fitted normal distributions for the distribution of the total activation energies of the spanning trees and spanning forests in Fig. \ref{fig:Histograms_and_Arrhenius} are: 
\begin{table}[h]
\begin{ruledtabular}
\begin{tabular}{cccc}
 Figure & $N_v$ & $E_\mathcal{T}$ $[\frac{\text{kJ}}{\text{mol}}]$& $E_\mathcal{F}$ $[\frac{\text{kJ}}{\text{mol}}]$\vspace{0.05cm}\\ \hline
 \ref{fig:Histograms_and_Arrhenius}(a,b) &$20$&$953.74 \pm 51.22$ &$914.26 \pm 58.23$\\\ref{fig:Histograms_and_Arrhenius}(d,e) &$100$&$4888.04 \pm179.66$ &$4845.02\pm 183.05$\\\ref{fig:Histograms_and_Arrhenius}(g,h) &$1.000$&$52756.76 \pm 607.74$ &$52696.46 \pm 618.11 $
\end{tabular}
\end{ruledtabular}
\caption{Fitting parameters for the normal distribution in Fig. \ref{fig:Histograms_and_Arrhenius}.}
\end{table}

Quadratic function of the shape $\ln{k}=aT^{-2}+bT^{-1}+c$ were fitted to the data in Fig \ref{fig:CrapseExamples}-\ref{fig:Powsner}. The corresponding parameters and the standard errors are listed in the following table:

\begin{table}[h]
\begin{ruledtabular}
\begin{tabular}{cccc}
 Figure & $a$ $[10^7K^2]$& $b$ $[10^5K]$& $c$ $[10^2\ln{\text{h}^{-1}}]$\\ \hline
 \ref{fig:CrapseExamples}(a) &$\textbf{-1.06}\pm 0.31$&$\textbf{0.68}\pm 0.21$&$\textbf{-1.10}\pm 0.36$\\
 \ref{fig:CrapseExamples}(b)& $\textbf{-1.81}\pm0.13$&$\textbf{1.19}\pm 0.09$&$\textbf{-1.98}\pm 0.15$\\
 \ref{fig:CrapseExamples}(c)& $\textbf{-1.96}\pm 0.14$&$\textbf{1.30}\pm 0.09$&$\textbf{-2.17}\pm 0.16$\\
 \ref{fig:CrapseExamples}(d)& $\textbf{-1.71}\pm 0.27$&$\textbf{1.13}\pm 0.19$&$\textbf{-1.88}\pm 0.31$\\
 \ref{fig:Bliss}(a)& $\textbf{-1.48}\pm 0.06$&$\textbf{0.97}\pm 0.04$&$\textbf{-1.60}\pm 0.07$\\
 \ref{fig:Bliss}(b)& $\textbf{-1.47}\pm 0.06$&$\textbf{0.97}\pm 0.04$&$\textbf{-1.59}\pm 0.07$\\
 \ref{fig:Powsner}(a)& $\textbf{-1.74}\pm 0.07$&$\textbf{1.14}\pm 0.04$&$\textbf{-1.88}\pm 0.08$\\
 \ref{fig:Powsner}(b)& $\textbf{-2.62}\pm0.18$&$\textbf{1.73}\pm 0.12$&$\textbf{-2.89}\pm 0.21$\\
 \ref{fig:Powsner}(c)& $\textbf{-2.36}\pm 0.22$ &$\textbf{1.56}\pm 0.15$&$\textbf{-2.59}\pm 0.25$\\
 \ref{fig:Powsner}(d)& $\textbf{-1.80}\pm0.04$&$\textbf{1.18}\pm 0.03$&$\textbf{-1.96}\pm 0.06$\\
 \ref{fig:Powsner}(e)& $\textbf{-1.75}\pm 0.07$&$\textbf{1.15}\pm0.04$&$\textbf{-1.91}\pm 0.08$\\
\end{tabular}
\end{ruledtabular}
\caption{Fitting parameters for the quadratic fit of the data from \cite{DevelopmentalArrhenius},\cite{Bliss} and \cite{Powsner} shown in the main text in Fig. \ref{fig:CrapseExamples}, Fig. \ref{fig:Bliss} and Fig. \ref{fig:Powsner}, respectively.}
\end{table}

\section{Generating the total activation energy distributions}
\label{app:DetailsSimulation}

The base graphs for the simulations in \ref{sec:LargeSystemLimit} were generated randomly, with the intent to display fundamental characteristics of a biochemical network involved in a developmental process. Specifically:
\begin{itemize}
    \item There must be a path from any vertex to the final vertex $N+1$.
    \item The distance from vertex $0$ to $N+1$ should be relatively large.
\end{itemize}
For these reasons, the standard Erdős–Rényi model, which creates edges with a fixed probability, was deemed inappropriate. Instead, the base graph was constructed and the total activation energies of the trees and two-tree spanning forests were generated using the following procedure:

\begin{enumerate}
    \item A function was defined to generate a random skeleton graph on $n$ vertices, organized into a hierarchical structure with the following properties:
    \begin{itemize}
        \item The first layer contains a single source vertex, and the final layer contains a single target vertex.
        \item The remaining vertices are assigned to intermediate layers with a probability $p^k$, where $k$ is the number of vertices already in the current layer. If not added to the current layer, the vertex is assigned to the next layer.
        \item Each vertex is connected by an outgoing edge to a randomly chosen vertex in the next layer and to an incoming edge from a randomly chosen vertex in the previous layer.
    \end{itemize}
    This protocol ensures that the first vertex acts as a source and the final vertex as a target.

    \item The function was then used to generate the main graph with $N$ vertices, divided into $n$ clusters:
    \begin{itemize}
        \item The $N$ vertices were randomly partitioned into $n$ non-empty sets.
        \item Each cluster was generated as a skeleton graph for the vertices in one partition set. To also include reversible steps within clusters, a back rate was added to each edge with a probability $q$.
        \item A skeleton graph on $n$ vertices was then generated to represent the interconnections between clusters. In this skeleton, each vertex was replaced by a cluster, ensuring that the source and target vertices of the cluster corresponded to the edges in the skeleton graph.
    \end{itemize}
    The resulting graph has a global source and a global target vertex that are relatively far apart. Additionally, it contains reversible steps within clusters and irreversible steps between clusters.
    \item For the reaction rates, an $N\times N$ matrix was generated, with entries $E_{ij}$ interpreted as activation energies for the $k_{ij}$ rates:
    \begin{itemize}
        \item     Entries above the main diagonal $(i<j)$ represent forward rates and entries below the main diagonal $(i>j)$ represent backward rates.
        \item The activation energies were drawn from uniform probability distributions. For the forward rates from  $[60 \frac{\text{kJ}}{\text{mol}},100 \frac{\text{kJ}}{\text{mol}}]$ and for the backward rates from $[20 \frac{\text{kJ}}{\text{mol}},60 \frac{\text{kJ}}{\text{mol}}]$.    
    \end{itemize}
    Uniform distributions were chosen as a parsimonious alternative to Gaussian-like distributions, reflecting the typical range of activation energies in biochemical reactions\cite{DevelopmentalArrhenius,DevelopmentalArrheniusRanges}.
    \item To study the behavior for large $N$, $100$ graphs were generated for $N+1=10$ to $N+1=100$ in steps of $10$, and for $N+1=200$ to $N+1=1.000$ in steps of $100$.
    \begin{itemize}
        \item For all cases, $n=N/10$ and $p=q=0.5$ were chosen.
        \item The distributions of total activation energies for spanning trees and two-tree spanning forests were computed.
        \item $10.000$ spanning trees and at least $10.000$ two-tree spanning forests were sampled for each case.
    \end{itemize}
\end{enumerate}
The results of these simulations are presented in the main text.

\section{Algorithm to compute the mean first-passage time in graph-theoretical framework}
\label{app:Algorithm}

To evaluate expressions like Eq. (\ref{eq:MFPTRatioOfZs}) for large networks, it is necessary to consider all spanning trees and two-tree spanning forests. As discussed in the main text, the number of such graphs grows very rapidly with the size of the network. Except in cases where the network has a simple, recurrent structure (such as the examples provided in Appendix \ref{app:MFPTCountingExample}), obtaining all these graphs quickly becomes impractical. Therefore, a probabilistic approach is more feasible to obtain a representative sample of the required spanning trees and two-tree spanning forests. Note that Eq. (\ref{eq:MFPTRatioOfZs}) is given by:

\begin{equation}
    \langle\tau\rangle =\frac{|\mathcal{F}|}{|\mathcal{T}|}\hspace{0.1cm}\frac{\frac{1}{|\mathcal{F}|}\sum_{\mathcal{F}}w(\mathcal{F})}{\frac{1}{|\mathcal{T}|}\sum_\mathcal{T}w(\mathcal{T})}=\frac{|\mathcal{F}|}{|\mathcal{T}|}\hspace{0.1cm} \frac{\langle w(\mathcal{F})\rangle _\mathcal{F}}{\langle w(\mathcal{T})\rangle _\mathcal{T}},\label{eq:MFPT_Algorithm}
\end{equation}
where $\sum_\mathcal{T}$ and $\sum_\mathcal{F}$ denote sums over all appropriate spanning trees and two-tree spanning forests, respectively. An algorithm to evaluate this expression thus has three objectives:
\begin{enumerate}
    \item Sample $\mathcal{T}$ at random and evaluate $\langle w(\mathcal{T})\rangle$
    \item Sample $\mathcal{F}$ at random and evaluate $\langle w(\mathcal{F})\rangle$
    \item Determine the ratio $\frac{|\mathcal{F}|}{|\mathcal{T}|}$
\end{enumerate}

\textbf{Objective 1:} Sampling random trees $\mathcal{T}$ rooted at vertex $N+1$ is a well-known problem in graph theory \cite{lyons2017probability}. The most efficient solution is Wilson’s algorithm \cite{wilson1996generating}, which is based on a loop-erased random walk.

\textbf{Objective 2:} Sampling random two-tree forests $\mathcal{F}$, where one tree is rooted at vertex $N+1$ and the other one is rooted at any another vertex and contains the initial vertex $1$, involves multiple steps to ensure that all possible forests are drawn with equal probability.

\begin{figure}[h]
    \centering
    \includegraphics[width=0.48\textwidth]{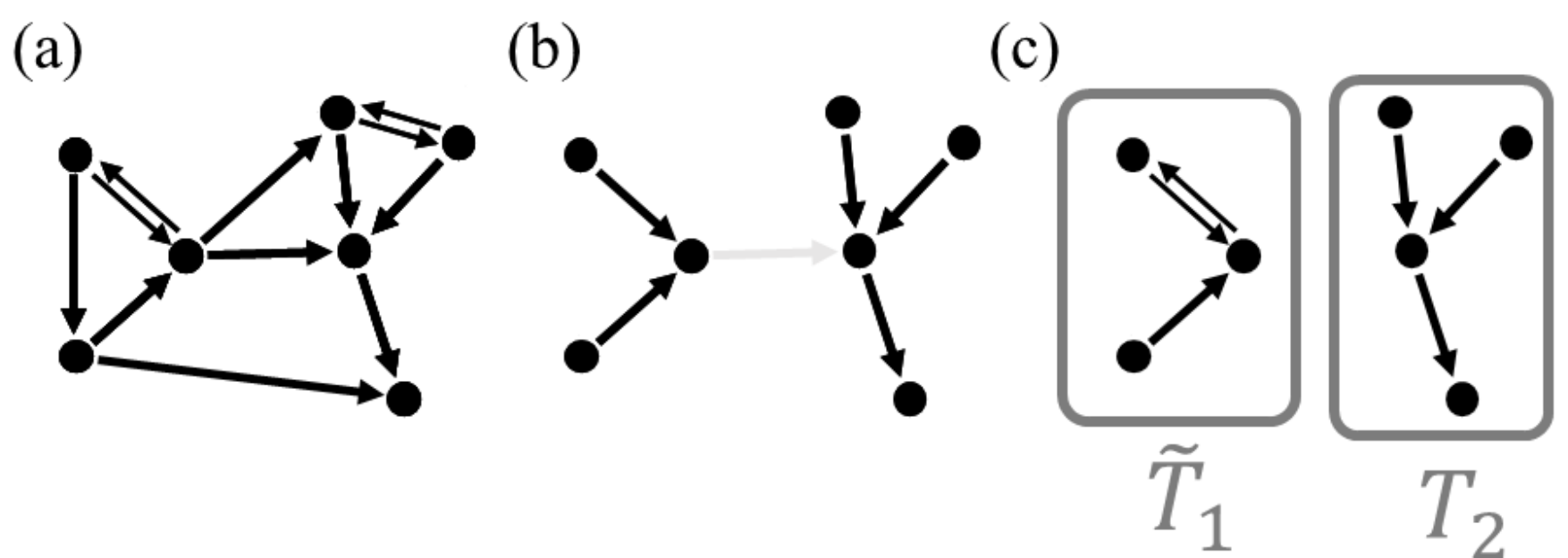}
     \caption{Part one of the algorithm. The first steps in the algorithm to get a random two-tree forests $\mathcal{F}$ of graph $G$: (a) an example graph $G$ (b) find a random spanning tree of $G$ and remove an edge from it to obtain two components (c) add to the tree without vertex $N+1$ any $(y,x)\in E(G)$, if $(x,y)$ is on the tree.}
     \label{fig:Algorithm_1}
\end{figure}

Start by drawing a random spanning tree $\mathcal{T}$ from the full graph (e.g., using Wilson’s algorithm) and remove a random edge $(r,v)$ from the path from $1$ to $N+1$ in $\mathcal{T}$. While the resulting graph is a two-tree spanning forest with $r$ and $N+1$ as roots of the two trees, this procedure does not yield all possible two-tree spanning forests. This is immediately clear when considering the example of a one-step master equation as discussed in Appendix \ref{app:MFPTCountingExample}: the (in this case unique) spanning tree does not contain any back rates, which are, however, present on most two-tree spanning forest. 
To take this into account,  for any edge $(x,y)$ on the tree without vertex $N+1$, $(y,x)$ has to be added to the tree, if  $(y,x)\in E(G)$. Call the resulting subgraph $\Tilde{T}_1$. The other component $T_2$ is a tree rooted at $N+1$ and will indeed be one of the two trees of the forest. These first steps are illustrated in Fig. \ref{fig:Algorithm_1}.

\begin{figure}[h]
     \centering
    \includegraphics[width=0.48\textwidth]{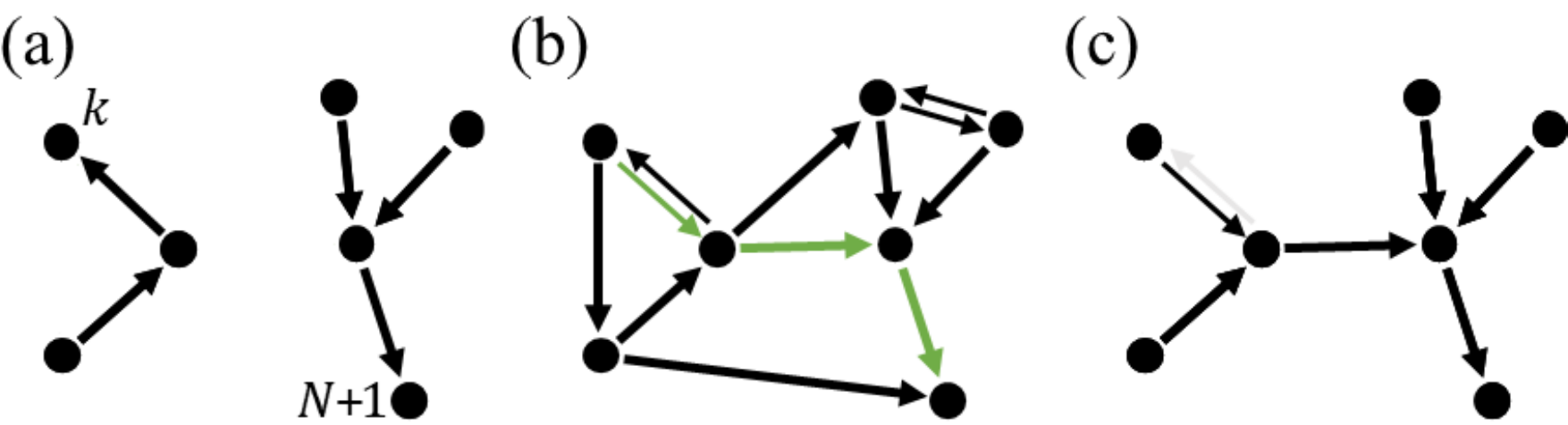}
     \caption{Part two of the algorithm. Illustration to show that any two-tree spanning forest is the subgraph of some graph constructed as in Fig. \ref{fig:Algorithm_1}: (a) choose any two-tree spanning forest (b) choose a path from $k$ to $N+1$ in $G$ (c) Add path to forest and remove opposite edges.}
     \label{fig:Algorithm_2}
\end{figure}

Any two-tree spanning forest $\mathcal{F}$ on $G$ is a subgraph of some $\Tilde{T}_1\cup T_2$ obtained this way. To see this, let $T^{'}_{k}$ and $T^{'}_{N+1}$ be the two trees in $\mathcal{F}$ with roots $k$ and $N+1$, respectively. Choose a path from $k$ to $N+1$ in $G$, follow it and add all its edges to  $T^{'}_{k}\cup T^{'}_{N+1}$ until the path reaches a vertex in $T^{'}_{N+1}$ for the first time. Remove any edges that point in opposite direction to the edges that were added. 
The resulting graph will then be a spanning tree of $G$. If $\Tilde{T}_1\cup T_2$ is constructed from it identifying the last edge added from the path as $(r,v)$, one has $T^{'}_{k}\subseteq\Tilde{T}_1$ and $T^{'}_{N+1}=T_2$. This construction is illustrated in Fig. \ref{fig:Algorithm_2}.

Next, identify all two-tree spanning forests in $\Tilde{T}_1\cup T_2$. This requires finding all vertices that can serve as roots for a spanning tree of  $\Tilde{T}_1$. These vertices are precisely those reachable from $r$ in $\Tilde{T}_1$, $R=\{r_0,...,r_m\}$ with $r_0:=r$, and can be found using basic breadth-first or depth-first search algorithms \cite{cormen2022introduction}. 
Construct the spanning tree $T_{1,r_i}\subseteq \Tilde{T}_1$ for each $i=0,...m$. Note that there is exactly one such tree for each $r_i$ because the graph obtained from turning all edges of $\Tilde{T}_1$ into undirected edges is a (undirected) tree and a given root uniquely defines its orientation.  

Now each $T_{1,r_i}\cup T_2$ yields a two-tree spanning forest and any two-tree spanning forest can be obtained from such a construction. But there is still one caveat: so far, these forests do not constitute a uniformly-drawn sample because some forests are more likely to be obtained than others.  

\begin{figure}[h]
     \centering
    \includegraphics[width=0.45\textwidth]{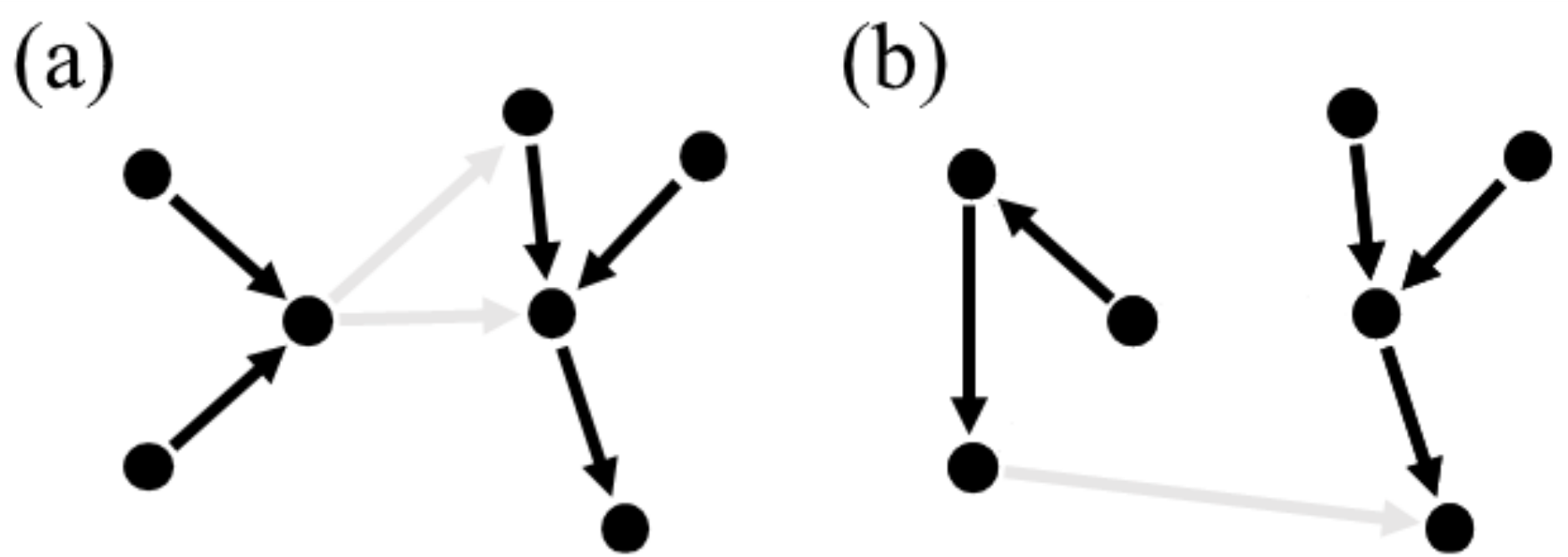}
     \caption{Part three of the algorithm. The initial partitions are degenerate; some subgraphs are induced by more trees than others: (a) the same subgraph is induced by removing an edge of two different spanning trees (b) a subgraph that can only be obtained by edge removal from a single spanning tree.}
     \label{fig:Algorithm_3}
\end{figure}

The first problem is that the same  $\Tilde{T}_1\cup T_2$ partition may be obtained from different spanning trees, as is shown in Fig. \ref{fig:Algorithm_3}. 

The degeneracy $d$ is precisely the number of directed edges from $R$ to $T_2$ and the probability of obtaining this partition is proportional to the degeneracy. To ensure the same probability for all partitions, one should accept a partition with probability $P(\text{"accept }\Tilde{T}_{1}\cup T_2\text{"})=d^{-1}$.

The second problem arises from the fact that the number of $r_i$ is not the same for different $\Tilde{T}_{1}$. Indeed, if only one $T_{1,r_i}\subseteq\Tilde{T}_{1}$ were chosen, the probability to choose $\Tilde{T}_{1}$ in the first place should be proportional to $|R|$. However, achieving this for general networks would require accepting it with a probability of $\frac{|R|}{N}$ - which is too small to be an efficient method for generating spanning forests. Since one is only interested in sampling the mean, it is more practical to  keep \textbf{all} the $T_{1,r_i}$ for $i=0,...m$ instead.   

\begin{figure}[h]
     \centering
    \includegraphics[width=0.45\textwidth]{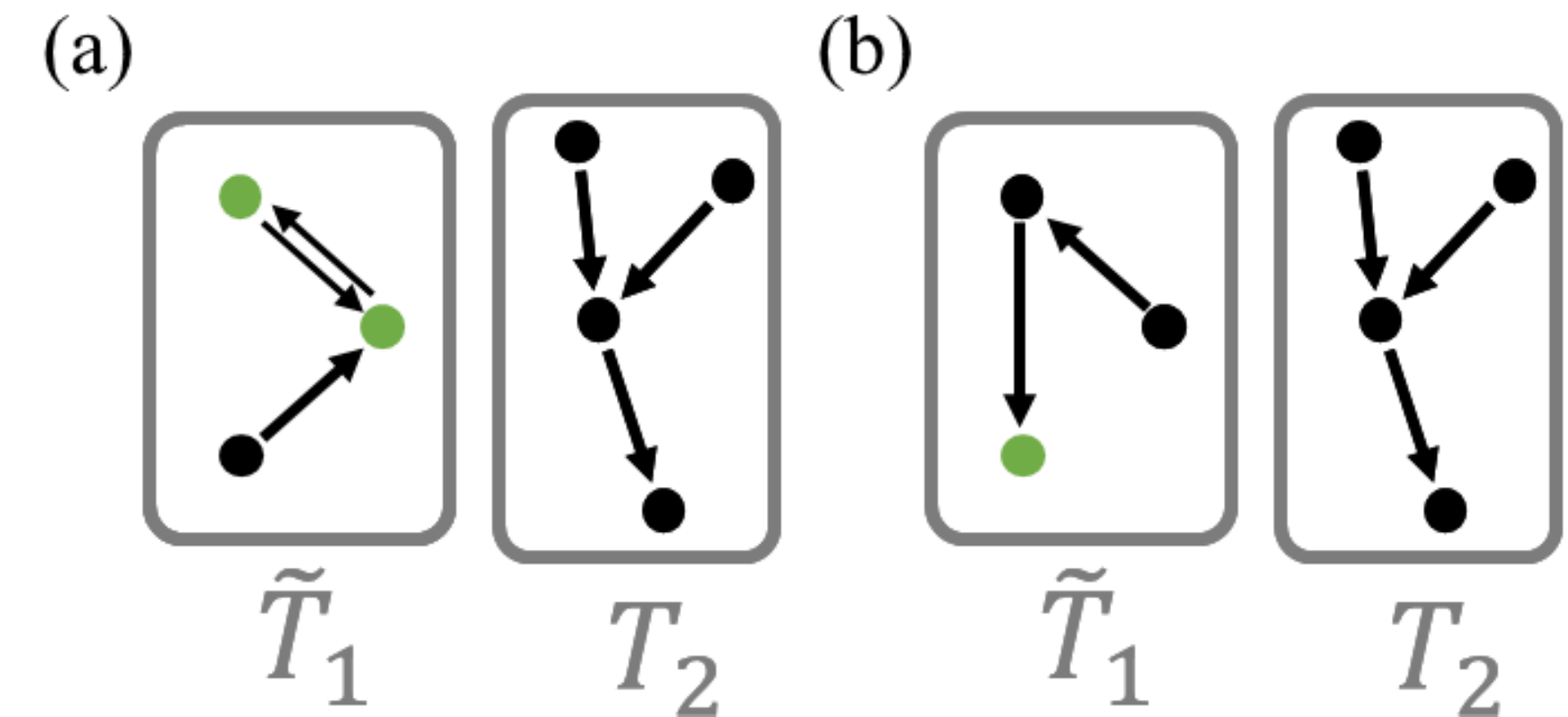}
     \caption{Part four of the algorithm. The partitions can contain a different number of potential roots, which has to be accommodated for to ensure a uniformly-drawn sample. (a) A partition where $\Tilde{T}_{1}$ contains two roots (green vertices) (b) A partition where $\Tilde{T}_{1}$ contains one root (green vertex).} 
     \label{fig:Algorithm_4}
\end{figure}

Note that if one only seeks to evaluate Eq. (\ref{eq:MFPT_Algorithm}), there is a third possibility that is even simpler: one may directly compute:
\begin{equation}
    \langle w(\mathcal{F})\rangle_\mathcal{F}=\frac{\sum_{m=1}^M \frac{|R_m|}{d_m}w(\mathcal{F}_m)}{\sum_{m=1}^M \frac{|R_m|}{d_m}},\label{eq:direct_estimate_w(F)}
\end{equation}
where $M$ is the number of two-tree forests that were drawn, $\mathcal{F}_m$ the $m$-th forest and $|R_m|$ and $d_m$ the number of its roots and its degeneracy, respectively.

The essential steps to get the two-tree spanning forests and the weights are summarized as a pseudocode (Algorithm \ref{alg:pseudo_code_algorithm}). 
\begin{algorithm}[h]
\caption{Find Random Spanning Forests of Two Trees of the Directed Graph \(G\)}
\label{alg:pseudo_code_algorithm}
\KwIn{A directed graph \(G\)}
\KwOut{Random two-tree spanning forest of \(G\), one tree rooted in \(N+1\) and the other one rooted at any other vertex together with the weight of this choice \(|R|/d\)}
$v_{\text{end}} \gets \max(\text{vertices of } G)$\;
    $T \gets \text{directed\_spanning\_tree}(G, v_{\text{end}})$\;
    \text{edges} $\gets \text{list of edges of } T$\;
    \text{random\_edge} $\gets$ \text{choose a random edge from edges}\;
    $T \gets T \setminus$ \text{random\_edge}\;
    $r_0 \gets$ \text{source vertex of random\_edge}\;
    $T_{\text{undirected}} \gets T.\text{to\_undirected}()$\;
    \text{C} $\gets$ \text{connected components of } $T_{\text{undirected}}$\;
    \text{trees} $\gets [T_{\text{undirected}}.\text{subgraph}(c).\text{copy()} \text{ for c in C}]$\;
    $T_1, T_2 \gets \text{trees}[0], \text{trees}[1]$\;
    \If{$v_{\text{end}} \in T_1.\text{vertices}()$}{
        $T_1, T_2 \gets T_2, T_1$\;
    }
    $T_{1,\text{directed}} \gets \text{inherit\_directions}(G, T_1)$\;
    $R \gets \text{descendants}(T_{1,\text{directed}}, r_0)$\;
    \text{d} $\gets 0$\;
    \For{$u \in T_2.\text{vertices}()$}{
        \For{$v \in \text{end\_vertices\_1}()$}{
            \If{$G.\text{has\_edge}(v, u)$}{
                \text{d} $\gets \text{d} + 1$\;
            }
        }
    }
    $r \gets \text{choose a random root from } R$\;
    $T_{1,\text{oriented}} \gets \text{direct\_towards\_root}(T_1, r)$\;
    $T_{2,\text{oriented}} \gets \text{direct\_towards\_root}(T_2, v_{\text{end}})$\;
    $F \gets \text{compose}(T_{1,\text{oriented}}, T_{2,\text{oriented}})$\;
    \Return $F, \frac{|R|}{d}$\;
\end{algorithm}

However, the objective of the simulations performed for this work was to obtain the probability distributions of the total activation energies along the two-tree spanning forests, which requires a uniformly drawn base sample. Therefore, the slightly more complicated and wasteful procedure explained before was used.

\textbf{Objective 3:} Finding $\frac{|\mathcal{F}|}{|\mathcal{T}|}$ can be integrated in the part where the two-tree spanning forests are generated. Note that any spanning tree has as many edges that can be removed, as there are edges on the path from $1$ to $N+1$. Once an edge is removed, there are $|R|$ two-tree spanning forests obtained from this tree. However, these are obtained from $d$ different spanning trees, and one has to correct for this again. In total, the ratio can be estimated as:
\begin{equation}
    \frac{|\mathcal{F}|}{|\mathcal{T}|} =\frac{1}{M}\sum_{m=1}^M L_m\frac{|R_m|}{d_m},
\end{equation}
where $L_m$ is the length of the path from $1$ to $N+1$ on the $m$-th tree.
Observe that Eq. (\ref{eq:direct_estimate_w(F)}) can now be written as:
\begin{equation}
    \langle w(\mathcal{F})\rangle_\mathcal{F}=\frac{1}{M}\frac{|\mathcal{T}|}{|\mathcal{F}|}\sum_{m=1}^M L_m\frac{|R_m|}{d_m}w(\mathcal{F}_m),
\end{equation}
and inserting this into Eq. (\ref{eq:MFPT_Algorithm}) yields:
\begin{equation}
    \langle\tau\rangle =\frac{1}{M} \frac{\sum_{m=1}^M L_m \frac{|R_m|}{d_m}w(\mathcal{F}_m)}{\langle w(\mathcal{T})\rangle _\mathcal{T}}.
\end{equation}
So, this method does not even require to explicitly calculate $\frac{|\mathcal{F}|}{|\mathcal{T}|}$.

\begin{figure*}[bth]
     \centering
    \includegraphics[width=1\textwidth]{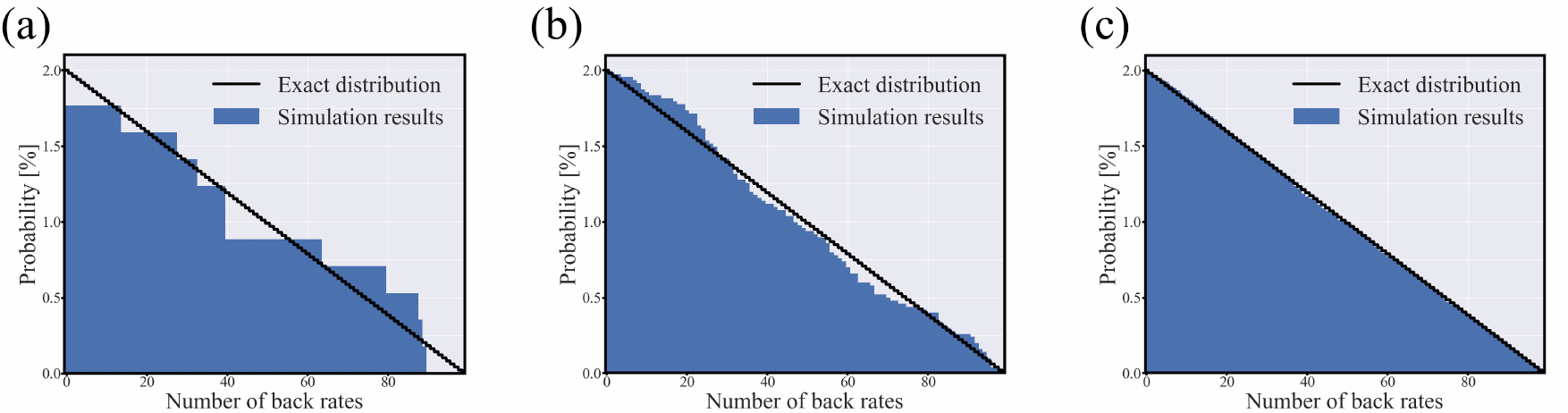}
     \caption{Distribution of the number of back rates on the two-tree spanning forests for samples of increasing size $n$ generated by the algorithm. The base graph is a linear graph on $100$ vertices with forward and backward rates (one-step master equation). The black curve shows the exact distribution from the analytical formula. (a) $n =10$ (b) $n=100$ (c) $n=1.000$.}
     \label{fig:One_Step_Convergence}
\end{figure*}
To illustrate that the sampling algorithm for the two-tree spanning forests works, it was applied to the one-step master equation from appendix \ref{app:MFPTCountingExample}.
The two-tree spanning forests were already characterized before. They consist of all choices of $j<N+1$ and $j\leq m <N-1$ corresponding to the forest with back rates from $m$ to $j$ and forward rates otherwise (compare also Fig. \ref{fig:OneStep}b). So for a fixed number of back rates $l=m-j$, there are exactly $N-l$ choices for $j$ and hence, $N-l$ corresponding forests. In total, this means there are $\sum_{l=0}^N(N-l)=\frac{N(N+1)}{2}$ forests. So, if one draws uniformly a two-tree spanning forest, the probability $p(l)$ that it has $l$ back rates is given by:
\begin{align}
    p(l) =\frac{2}{N(N+1)}(N-l).
\end{align}
The exact distribution for $p(l)$ was compared to the one obtained by sampling with the algorithm for a one-step master equation on $100$ vertices, and one observes that the simulated distribution indeed converges to the exact solution, confirming that the algorithm yields a uniform sample (see Fig. \ref{fig:One_Step_Convergence})

\begin{figure*}[thb]
    \includegraphics[width=0.9\textwidth]{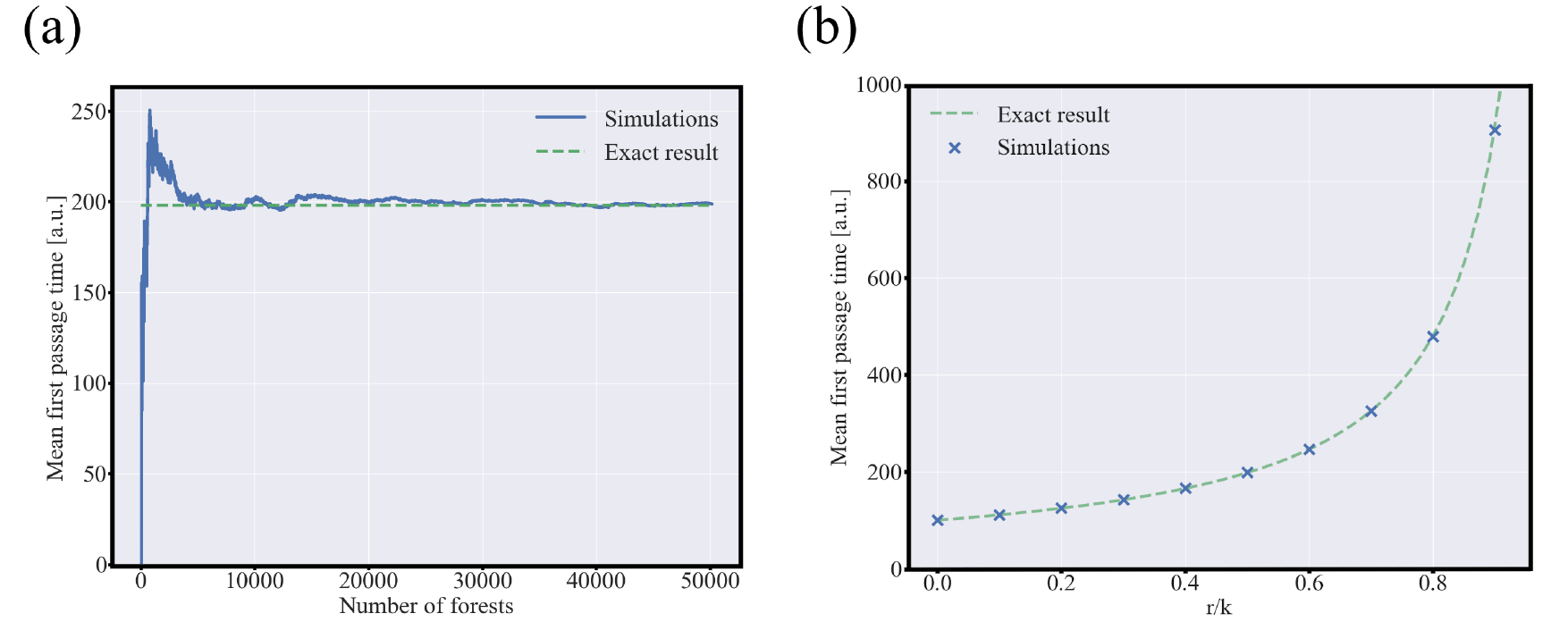}
    \caption{The algorithm applied to the one-step master equation. (a) MFPT obtained by the algorithm for an increasing number of two-tree spanning forests compared to the theoretical value for a one-step master equation on $100$ vertices with forward rates $k_i=1$ and backward rates $r_i=0.5$ for all $i$. (b) MFPT from same sample of two-tree spanning forests for different values of the back rate $r$ compared to the exact result.}
    \label{fig:One_Step_MFPT_algorithm}
\end{figure*}

The convergence of the MFPT obtained via the algorithm to the correct value is shown in Fig. \ref{fig:One_Step_MFPT_algorithm}a. The exact value is given by Eq. (\ref{eq:MFPTOneStep}).

An advantage of this algorithm over simulating the stochastic dynamics of the network is that it allows a sample of spanning trees and two-tree spanning forests to be generated once, enabling the computation of the mean first-passage time for different rates without regenerating the sample. This facilitates studying the impact of changing rates without the need to redo the entire simulation each time. To illustrate this, one sample of spanning forests of 50,176 spanning forests (from $1.000$ draws) was generated and the values of the MFPT were calculated for $k_i=1$ from $r_i=0$ to $r_i=0.9$ for all $i$ . The result is again compared to the exact solution given in Eq. (\ref{eq:MFPTOneStep}). This is shown in Fig. \ref{fig:One_Step_MFPT_algorithm}b.

\newpage
\vspace{2cm}


\providecommand{\noopsort}[1]{}\providecommand{\singleletter}[1]{#1}%

\end{document}